# Atomically thin p-n junctions based on two-dimensional materials

Riccardo Frisenda*[1], Aday J. Molina-Mendoza[2], Thomas Mueller[2], Andres Castellanos-Gomez*[3] and Herre S. J. van der Zant*[1,4,5]

[1] Instituto Madrileño de Estudios Avanzados en Nanociencia (IMDEA-Nanociencia), Campus de Cantoblanco, E-28049 Madrid, Spain.

[2] Institute of Photonics, Vienna University of Technology, Guhausstrae 27-29, 1040 Vienna, Austria.

[3] Materials Science Factory, Instituto de Ciencia de Materiales de Madrid (ICMM-CSIC), E-28049, Madrid, Spain.

[4] Kavli Institute of Nanoscience, Delft University of Technology, Lorentzweg 1, 2628 CJ Delft, The Netherlands.

[5] Departamento de Fisica de la Materia Condensada, Universidad Autonoma de Madrid, Campus de Cantoblanco, E-28049 Madrid, Spain.

*E-mail: riccardo.frisenda@imdea.org, andres.castellanos@csic.es, h.s.j.vanderzant@tudelft.nl

**ABSTRACT:** Recent research in two-dimensional (2D) materials has boosted a renovated interest in the p-n junction, one of the oldest electrical components which can be used in electronics and optoelectronics. 2D materials offer remarkable flexibility to design novel p-n junction device architectures, not possible with conventional bulk semiconductors. In this Review we thoroughly describe the different 2D p-n junction geometries studied so far, focusing on vertical (out-of-plane) and lateral (in-plane) 2D junctions and on mixed-dimensional junctions. We discuss the assembly methods developed to fabricate 2D p-n junctions making a distinction between top-down and bottom-up approaches. We also revise the literature studying the different applications of these atomically thin p-n junctions in electronic and optoelectronic devices. We discuss experiments on 2D p-n junctions used as current rectifiers, photodetectors, solar cells and light emitting devices. The important electronics and optoelectronics parameters of the discussed devices are listed in a table to facilitate their comparison. We conclude the Review with a critical discussion about the future outlook and challenges of this incipient research field.

# 1. Introduction

The p-n junction has become an important component in modern electronics since its serendipitous discovery by Russel Ohl almost 80 years ago.[1] This kind of device can be created by joining together two semiconductors of different type: a p-type semiconductor containing an excess of holes and n-type one with an excess of electrons. As a result an intrinsic electric field at the interface between them is generated and it can be used to rectify currents or to separate photogenerated electron-hole pairs. In bulk semiconductors the typical way to create a p-n junction is to dope two different parts of a single crystal with different ions or dopants forming a three-dimensional (3D) p-n junction. The resulting face to face arrangement constitutes the only possible device architecture considered for bulk semiconducting materials.

If one scales down the dimensions of the semiconductors involved, by passing from 3D to two-dimensional (2D) materials, new and exciting possibilities arise. The design of p-n junctions offers now more possibilities and freedom: p-n junctions in the 2D case can be constructed following two main architectures: a lateral junction, in which the two 2D materials are joined at the same plane (creating a one-dimensional interface between the two materials) and a vertical junction, in which the 2D materials are stacked face to face, thereby exhibiting a two-dimensional overlap. Moreover, the ultra-thin nature of 2D materials gives rise to novel properties compared to 3D semiconductors.[2-8] An example is the thickness-dependency of the bandgap for some materials which enables even more possibilities to create different p-n junctions concepts.

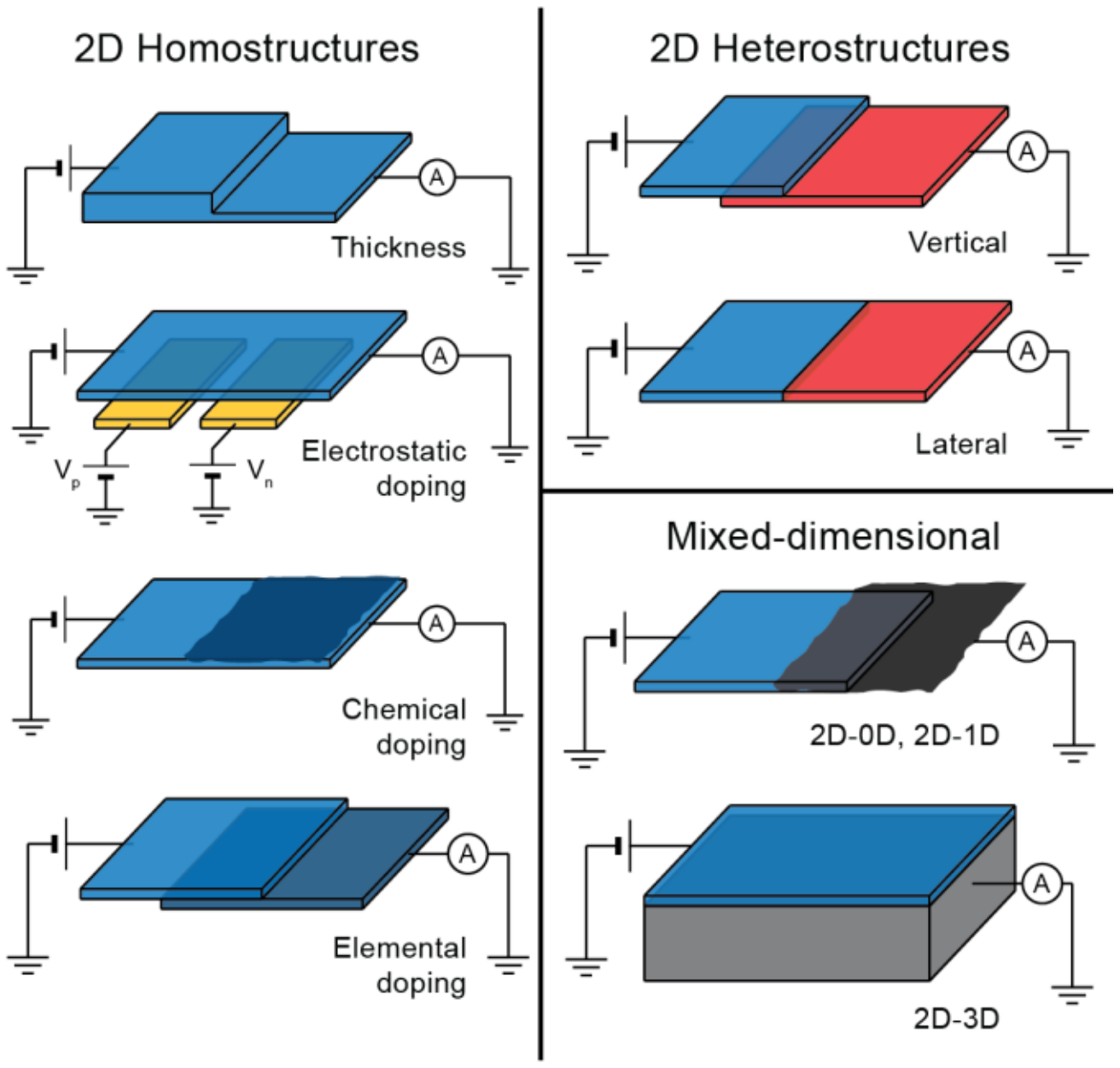

**Figure 1. Scheme of the different p-n junctions architectures based on 2D semiconductors.** Using two-dimensional (2D) materials a large variety of p-n junctions can be produced. These can be junctions based on a single 2D material (homostructures), on the junction between two different 2D materials (heterostructures) or junctions based on the combination of a 2D material with a material with higher or lower dimensionality (mixed-dimensional structures). Different concepts of p-n junctions within these categories are shown.

Figure 1 shows a schematic of eight kinds of p-n junctions based on 2D materials which one can encounter in the present literature. We separate homojunctions (based on a single 2D material), from heterojunctions (formed by joining two different 2D materials) and mixed dimensional junctions (based on the combination of a 2D material with 0D, 1D or 3D materials. More specifically, the eight kinds of p-n junctions are:

*2D homostructures*

1. Thickness-based junctions, in which the p- and n-regions are formed by two regions of the same material with different thicknesses.
2. Electrostatically doped junctions, in which the doping in different regions of the same 2D material is controlled by local electrostatic gates.
3. Chemical doping, in which the doping of a region in a 2D material is modified by the adsorption of molecules, nanoparticles or quantum dots onto the surface of the material.
4. Elemental doping, in which two flakes of the same 2D material with different doping are stacked one on top of the other, forming an out-of-plane junction.

*2D heterostructures*

5. Vertical heterojunctions, in which two different 2D materials are stacked one on top of the other and the junction is formed in the out-of-plane direction.
6. Lateral heterojunctions, in which two 2D materials are joined in the same plane along a one-dimensional interface.

*Mixed-dimensional*

7. 2D-0D and 2D-1D p-n junctions, in which a molecular crystal or a nanotube film is in contact with a 2D material.
8. 2D-3D p-n junctions, in which a 2D material is in contact with a bulk 3D semiconductor.

## 2. Semiconducting 2D materials building blocks and tools

The different p-n junction architectures introduced in Section 1 can be realized thanks to the discovery of 2D materials. In this section we will describe the atomic and electronic structure of some of the main 2D materials which can be used to fabricate p-n junctions. We will then introduce the production and isolation methods for these materials discussing top-down approaches such as the deterministic transfer and bottom-up methods such as CVD growth.

### 2.1. Atomic and electronic structure

2D materials can be extracted out of layered materials by mechanical or chemical exfoliation.[9-12] The atoms in these materials are arranged in layers with strong in-plane bonds (typically covalent bonds) and weak bonds between the different layers (generally van der Waals interactions). The layers can be composed of a single atomic plane, like in the case of graphene or hexagonal boron nitride (h-BN), or they can be made out of multiple atomic planes (e.g. in $MoS_2$ each monolayer is composed of three atomic layers). Many families of layered materials are known and in the following we will discuss the properties of some important examples of 2D materials.[10, 13-15]

Figure 2a shows the crystal structure of a monolayer of graphene, $MoS_2$ and h-BN. Graphene and h-BN share the same hexagonal crystal structure but differ strongly in their electronic structure. Graphene is a zero-gap semiconductor with a linear dispersion close to its neutrality point.[16] The doping in graphene can be controlled electrostatically and both n-type and p-type can be realized. Among the other methods used to dope graphene one can find chemical doping, substitutional doping and irradiation methods. Moreover many groups dedicated efforts to open a bandgap in graphene using external electric fields, geometric confinement or hydrogen adsorption.[17-21] In contrast, h-BN is an insulator with a bandgap of ~6 eV that makes it highly transparent to visible light.[22]

Among the semiconducting 2D materials the most studied ones are probably $MoS_2$ and other members of the transition metal dichalcogenide (TMDC) family that are composed of a layer with the transition metal atoms sandwiched between two layers of chalcogenides atoms.[2, 23-29] $MoS_2$ is a semiconductor with an indirect bandgap of 1.2 eV in its bulk or multilayer form but it exhibits a direct bandgap of 1.9 eV when thinned down to a monolayer.[30-32] Monolayer $WS_2$, $MoSe_2$ and $WSe_2$ share similar electronic properties, having direct bandgaps of energy between 1 and 2 eV. While $MoS_2$ and $WS_2$ are typically n-doped, $WSe_2$ is ambipolar. Another semiconducting 2D material largely used in p-n junctions is black phosphorous (BP). This layered allotrope of phosphorous is a narrow-bandgap semiconductor with a bandgap of 0.34 eV in bulk that increases to 1.5 eV when BP is thinned one monolayer thick.[33-35] Among the materials described BP is by far the most unstable as it degrades quickly when exposed to air and thus it requires to work in a high-vacuum environment or to employ encapsulation protocols (i.e., sandwiching the 2D material in between two protecting insulating layers) to prevent its degradation.[36-40] In addition to BP and graphene, the elemental (Xenes) family comprises also other materials such as silicene, germanene and stanene.[41, 42] Other important families of semiconducting layered materials are monochalcogenides such as SnS or GeSe,[43, 44] trichalcogenides such as $TiS_3$ or $HfS_3$[45, 46] and metal halides such as $PbI_2$ or $CrI_3$.[47, 48] Layered oxides, nitrides and carbides are also actively researched families of 2D materials.[49-51] Table 1 reports some of the important material properties of selected 2D materials useful for the construction of p-n junctions.

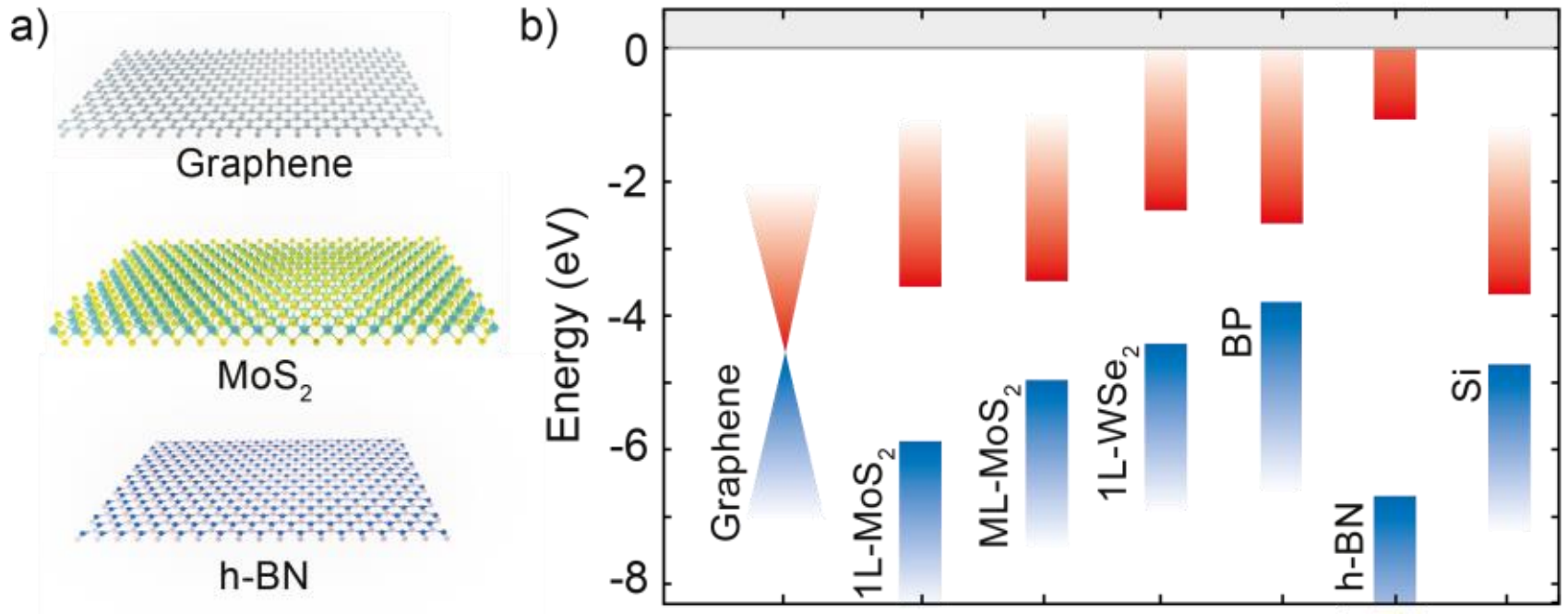

**Figure 2. Two dimensional materials employed as building blocks to fabricate p-n junctions.** a) A large number of two-dimensional (2D) layered materials with varying chemical composition, atomic structures and electronic properties are available. Starting from the top a zero-gap semiconductor, graphene, a semiconductor, $MoS_2$, and an insulator, h-BN are displayed. b) Schematic band diagram of important materials for 2D based p-n junctions showing the approximate band alignment between the different 2D materials according to the available literature values of the work function and band gap.

| Material | Bandgap (eV) | Electron affinity (eV) | Doping type |
|---|---|---|---|
| 1L-$MoS_2$ | 1.89 | 4.3 | n |
| ML-$MoS_2$ | 1.2 | 4.0 | n |
| 1L-$MoSe_2$ | 1.64 | 3.9 | n |
| 1L-$WS_2$ | 1.96 | 3.9 | n |
| 1L-$WSe_2$ | 1.6 | 3.6 | Ambipolar |
| 1L-BP | 1.5 | 3.9 | Ambipolar |
| ML-BP | 0.35 | 4.1 | Ambipolar |

**Table 1.** Bandgap energy, electron affinity and doping type of some layered materials. Bandgap energies and electron affinities are taken from Refs. [52-55].

## 2.2. Isolation of 2D materials

While 2D materials supported on bulk substrates are known since at least 50 years, their isolation and integration in devices is more recent. In 2004 Geim and Novoselov [9] demonstrated the first electronic device based on a 2D material, fabricated by isolating single-layer graphene from bulk graphite using mechanical cleaving. In this section we will discuss the different approaches developed to produce 2D materials and to integrate them in p-n junctions, including top-down approaches (which are very powerful in creating different and novel p-n junctions) and bottom-up approaches suitable for large-scale integration.

### 2.2.1. Top-down approach: deterministic transfer

Atomically thin 2D materials can be isolated from bulk layered crystals by mechanical cleaving. The presence of strong bonds in the plane of the layers and weak bonds between them allows these materials to cleave perfectly along the atomic plane. As mentioned above, the first technique developed is mechanical cleavage, also known as the "Scotch tape method". In this method a piece of tape is pressed against the surface of a natural or artificial bulk layered crystal and then it is peeled off, cleaving the crystal and leaving debris and flakes of the material onto the tape surface. The flakes can then be transferred to an arbitrary surface by pressing on it the tape containing the flakes and subsequently releasing the tape gently.
The main drawback of the Scotch tape method is that it produces flakes with different sizes and thicknesses randomly distributed over the sample substrate and only a small fraction of these flakes are atomically thin. This limitation can partially be circumvented by the use of optical identification methods to find atomically thin crystals from the crowd of thicker, bulky flakes.[56-58] Nevertheless, the combination of the Scotch tape method with optical identification methods alone cannot provide a reliable way to fabricate p-n junctions by artificial stacking of 2D crystals. Since 2010, the research on those artificial stacks of 2D materials has grown exponentially driven by the development of different transfer techniques that allow to place 2D materials on a desired location with an unprecedented degree of control and accuracy.[59-63] A typical deterministic placement setup used to transfer and artificially stack 2D crystals is schematically depicted in Fig. 3a. Such a setup is usually based either on a zoom lens or on an optical microscope equipped with long working distance objectives. Two manually actuated micropositioners are used to move the target surface and the flake to be transferred. Fig. 3b shows the steps to fabricate a vertical heterojunction using the deterministic transfer method. Figure 3c shows an optical image of a vertical homojunction formed by a n-doped $MoS_2$ flake stacked onto a p-doped $MoS_2$ one. The top image shows an intermediate fabrication step and the bottom image is the final device.

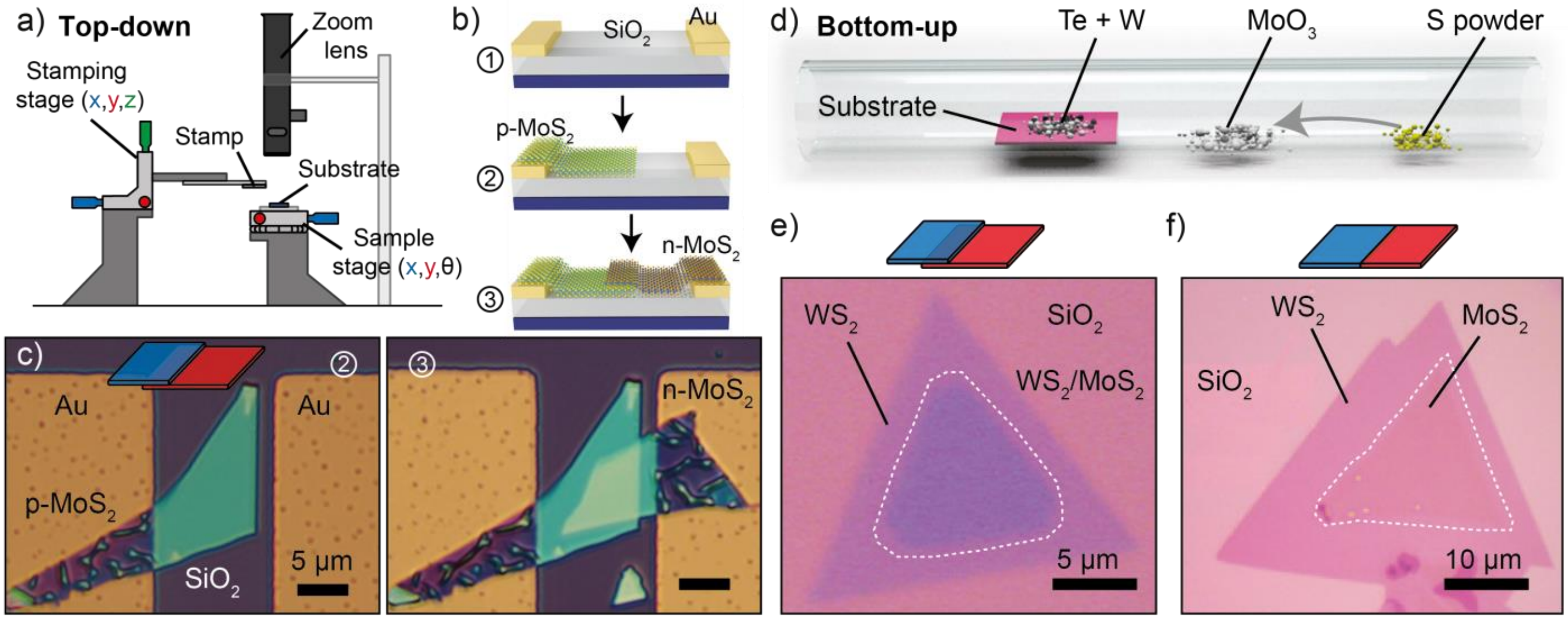

**Figure 3. Top-down and bottom-up approaches to fabricate p-n junctions.** a) Schematic diagram of a setup employed for the all-dry transfer of 2D crystals. b) Schematic diagram of the fabrication of a vertical heterojunction by a top-down approach. c) Sequential optical images of the fabrication of a p-n junction by vertical stacking of a p-type $MoS_2$ flake onto a n-type $MoS_2$ one. d) Schematic diagram of the synthesis process of $MoS_2/WS_2$ vertical or lateral heterostructures with a bottom-up method. e) Optical image of a vertically stacked $WS_2/MoS_2$ heterojunction synthesized at 850 °C. f) Optical image of a lateral $WS_2/MoS_2$ heterojunction grown at 650 °C. Panel (c) readapted from Ref. [64] with permission from John Wiley and Sons and panels (d), (e) and (f) readapted from Ref. [65] with permission from Springer Nature.

### 2.2.2. Bottom-up approach: CVD growth

Many 2D materials can be directly grown in the form of single- or few-layer nanosheets on various substrates.[66] The production process usually involves a thermal chemical vapour deposition (CVD) process, in which the vapour-phase reactants are generated by thermally evaporation of specific source materials.[67-69] This growth method is very promising for the production and large-scale integration of CVD grown 2D p-n junctions and already many vertically stacked and lateral p-n junctions have been demonstrated. The family of TMDCs is particularly interesting for CVD growth because of the relatively small lattice mismatches between its various members and the availability of different doping types.

In 2014 Gong *et al.*[65] demonstrated a single-step vapour phase growth process for the creation of lateral or vertical monolayer $MoS_2/WS_2$ heterostructures controlled by the growth temperature. Figure 3d shows a schematic of the synthesis, which involves sulphur powder, $MoO_3$ as a source of molybdenum and a mixed powder of W and Te for the tungsten. The difference in nucleation and growth rates between $MoS_2$ and $WS_2$ allows for the sequential growth of $MoS_2$ and $WS_2$ instead of the formation of a $Mo_xW_{1-x}S_2$ alloy. By carrying out the process at a temperature of ~850 °C the authors produced vertically stacked $WS_2$-$MoS_2$ bilayers, similar to the structure shown in Fig. 3e, whereas at ~650 °C in-plane lateral heterojunctions were created shown in Fig. 3f. In another study Duan *et al* [70] demonstrated the growth of lateral $MoS_2$-$MoSe_2$ and $WS_2$-$WSe_2$ heterojunctions by switching *in situ* the vapour-phase reactants to enable lateral epitaxial growth of single- or few-layer TMDC heterostructures. Finally, the growth of $MoSe_2$-$WSe_2$ lateral heterojunctions was demonstrated by Huang and coauthors.[71] Recently a different approach was demonstrated by Zheng *et al.*, which used pulsed laser deposition to achieve a single-step growth of a lateral p-n junction between layered $In_2Se_3$ and nonlayered $CuInSe_2$.[72] The biggest challenge in CVD process of heterojunctions is probably the interplay between the many degrees of freedom of the system (such as temperature, flow rate, substrate, lattice mismatch and others) which makes it difficult to ensure the reproducibility of the final process outcome in different laboratories and therefore it requires the full optimization of the growth recipe in every new growth setup.

## 3. p-n junctions based on 2D materials

Using to the top-down and bottom-up approaches described in the previous section, a large number of p-n junction devices based on 2D materials has been demonstrated in literature. In this section, we will review these results by passing from homojunctions to heterojunctions and ending with the mixed-dimensional junctions.

### 3.1. Homostructures

Homojunctions devices are p-n junctions based on a single 2D material. Figure 4a-c show three examples of such junctions: (1) based on quantum-confinement effects, (2) electrostatic gating and (3) on chemical doping to obtain a spatial variation of the doping profile. A fourth kind of homojunction is based on elemental doping of 2D materials.

#### 3.1.1. Thickness modulation

Figure 4a shows a 2D homojunction based on thickness modulation. In ultra-thin materials, the bandgap energy becomes a thickness-dependent quantity because of quantum-confinement effects. This permits the creation of a p-n junction in which the p and n regions are made of the same material, just with different thickness. One example shown in Fig. 4d is a p-n junction based on a single $WSe_2$ flake.[73] To fabricate this device Xu and coauthors started from a bilayer $WSe_2$ flake and then partially thinned the flake to a monolayer with an Ar plasma. Metallic contacts to the monolayer and bilayer regions were subsequently defined. To fabricate this kind of homojunctions one can also take advantage of the exfoliation process itself as it generally produces flakes that are already composed of different thickness regions.[74]

### 3.1.2. Electrostatic doping

The reduced thickness of 2D materials usually comes along with a large electric field effect tunability. Figure 4b shows a schematic diagram of an electrostatically defined 2D p-n junction and Fig. 4e shows the optical picture of the actual device. In this case $WSe_2$ is used because of its ambipolar nature (by tuning the polarity of the gate voltage one can make it either n- or p-type). Several groups reported this kind of device.[75-79] Here, we will discuss the implementation by Pospischil and coauthors. In their work, two metallic gates separated by 460 nm are defined onto a $SiO_2$/Si substrate and subsequently covered by a 100 nm thick $Si_3N_4$ gate dielectric. A flake of mechanically exfoliated $WSe_2$ was then deterministically transferred, partly covering both prepatterned gates electrodes, and source and drain electrodes were defined afterwards. By independently controlling the two split gates the doping in the two regions of the $WSe_2$ flake located above the local gates can be controlled. Similar devices based on this split-gate geometry were fabricated with BP[80] and graphene.[81, 82] A different geometry was recently used by Li *et al.*[83] in which a single local graphene gate electrode, insulated by a thin h-BN flake, is partly covered by a $WSe_2$ flake forming the active channel connected to source and drain electrodes. A complementary approach is to use ionic gating of the 2D material instead of the solid state split gates.[84-87]

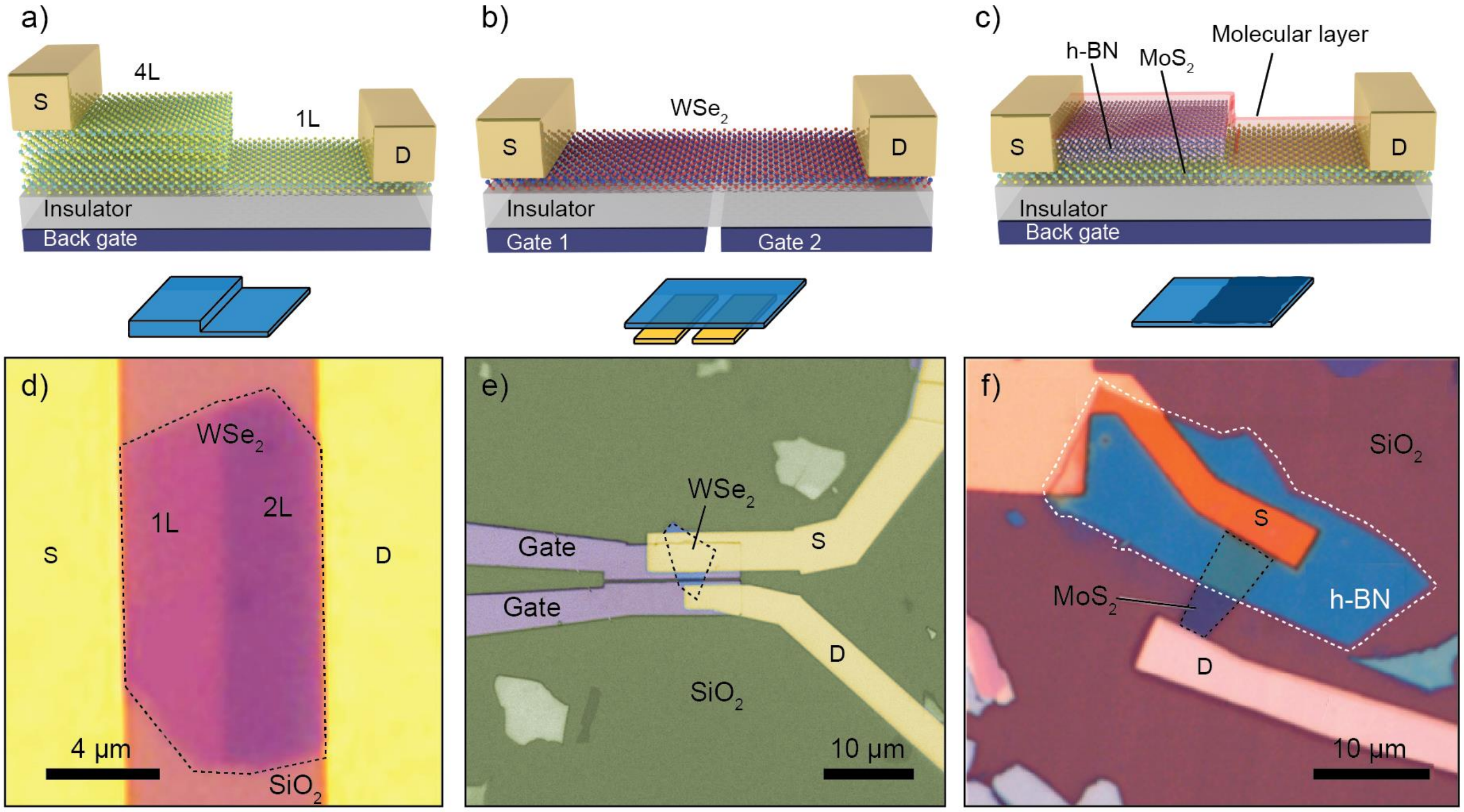

**Figure 4. Different examples of p-n junctions based on 2D materials homojunctions.** a-c) Schematic diagram of a p-n junction based on (a) $MoS_2$ with different thickness, (b) ambipolar $WSe_2$ with two split gate

electrodes and (c) local molecular doping of partially h-BN encapsulated $MoS_2$. d) Optical image of a thickness-modulated p-n junction based on single-layer and bilayer $WSe_2$. e) Optical image of a split-gate homojunction as schematically drawn in panel (b). f) Optical image of a chemically doped $MoS_2$ p-n junction as schematically illustrated in panel (c). The device is covered with molecules and part of the $MoS_2$ is protected by a h-BN flake. Panel (d) readaptedfrom Ref. [73] with permission from IOP Publishing, panel (e) readapted from Ref. [77] with permission from Springer Nature and panel (f) readapted from Ref. [88] with permission from American Chemical Society (copyright 2014).

#### 3.1.3. Chemical doping

In addition to making 2D materials sensitive to external electric fields, the reduced thickness gives them a high sensitivity to the environment surrounding their surface. Molecules physisorbed or chemisorbed onto the surface can influence transport in the 2D material for example by introducing doping effects[89-94] and a p-n junction can thus be fabricated by locally doping the material. This principle has been used to fabricate chemically doped p-n junctions with graphene,[95, 96] TMDCs and black phosphorous.[97, 98] A chemically doped $MoS_2$ p-n junction is shown in Fig. 4f reported by Choi and coauthors.[88] In this work a solution of $AuCl_3$ was deposited onto the surface of a $MoS_2$ flake partially covered by an h-BN flake. $AuCl_3$ is known to induce p-type doping in $MoS_2$ thereby creating a lateral p-n junction, where the n region is provided by the natural n-doped $MoS_2$. On the other hand, a common molecule used to induce strong n-doping in $MoS_2$ is benzene viologen (BV). Li and coauthors[99] used BV and $AuCl_3$ to create an out-of-plane p-n junction (a device in which charge transport is out-of-plane) where one face of a few layer $MoS_2$ flake was p-doped with $AuCl_3$ and the other one with BV.

#### 3.1.4. Elemental doping

A different approach to dope 2D materials is elemental doping, a technique that has proven to be successful for controlling carrier types in bulk materials. In bulk TMDCs elemental doping with Nb (five valence electrons), Fe or Re (seven valence electrons) has been used as a substitutional p-type or n-type dopant to replace the metallic atoms such as Mo or W (six valence electrons).[100] For 2D layered materials this technique was initially demonstrated for $MoS_2$ by Suh and coauthors.[101] Substitutional doping with Nb was used to convert $MoS_2$ from n-type to p-type and a vertical p-n homojunction was fabricated by stacking Nb doped $MoS_2$ onto undoped $MoS_2$. Similarly, Jin *et al.*[102] demonstrated a p-n homojunction by stacking undoped $MoSe_2$ onto doped $MoSe_2$ with Nb atoms. Figure 3b shows an optical image of a similar device fabricated by stacking p-type $MoS_2$ doped with Nb onto n-type $MoS_2$ doped with Fe atoms.[64, 103] Recently, in a different approach phosphorous atoms were used to substitute the surficial sulfur atoms of $MoS_2$, leading to a p-type doping of $MoS_2$.[104, 105]

### 3.2. Heterostructures

The combination of two different 2D materials to form a heterostructure is one of the most promising strategies because of the large variety of bandgap energies and doping types available among the different 2D materials. Figure 5a-b show schematic diagrams of a vertical and a lateral heterojunction.

### 3.2.1. Vertical junctions

Vertical heterostructures represent a popular architecture in 2D materials. Due to a surface free of dangling-bonds and interlayer van der Waals interactions it is possible to stack 2D materials on top of each other without constraint on the lattice constants.[6, 106-108] The first example of a van der Waals heterostructure in literature was demonstrated by Dean *et al.*[60] that stacked graphene on top of an h-BN flake. Soon thereafter it was realized that by stacking an n-type 2D material onto a p-type 2D material, using the deterministic transfer method, atomically thin p-n junctions could be created.[109, 110] An example of an atomically thin vertical p-n junction[111] based on single-layer $MoS_2$ and $WSe_2$ is shown in Fig. 5c. The authors fabricated the heterojunction by deterministic placement of individual mechanically exfoliated monolayers of both materials on a $SiO_2$/Si substrate and deposited Pd contacts to inject electrons and holes into the n-$MoS_2$ and p-$WSe_2$ layers, respectively. In another work Lee and coauthors[112] introduced graphene contacts for the $MoS_2$ and $WSe_2$ monolayers to improve the collection of charges, realizing a graphene sandwiched 2D p-n junction. The freedom of stacking 2D materials on top of each other permitted the fabrication of many different vertical stacks by this top-down approach. Among the different devices we find junctions between two TMDCs, such as $WSe_2$-$MoSe_2$ or $MoTe_2$-$MoS_2$, and junctions of a TMDC with a different 2D layered material, BP-$MoS_2$, GaTe-$MoS_2$ or with ultrathin membranes such as InAs-$WSe_2$. Figure 5d shows a BP-$MoS_2$ vertical p-n junction[113] fabricated by transferring a flake of few-layer BP onto a monolayer $MoS_2$ grown by CVD onto a $SiO_2$/Si substrate. More recently vertical p-n junctions were directly fabricated by epitaxial growth without the need of mechanical assembly.

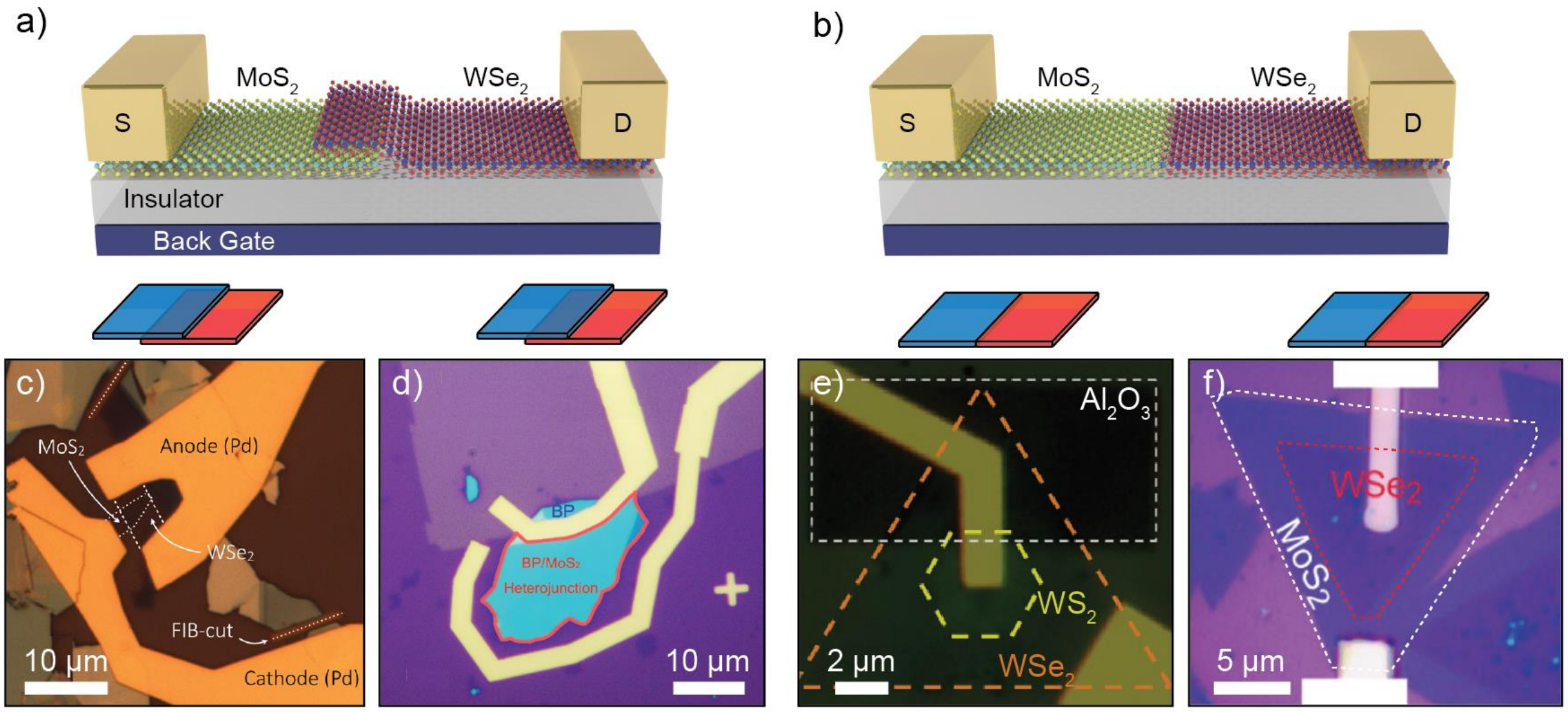

**Figure 5. P-n junctions based on heterojunctions between two different 2D materials.** a-b) Schematic diagram of a vertical (a) and lateral (b) p-n junction. c) Optical image of a vertical p-n junction formed by stacking a monolayer $MoS_2$ flake onto a monolayer $WSe_2$ flake. d) Optical image of a vertical $MoS_2$-black

phosphorous device. The dark purple region is monolayer $MoS_2$, while the blue flake is few-layer black phosphorus. The light purple region is $SiO_2$. e) Optical image of a lateral $WSe_2$-$WS_2$ p-n junction. The white dashed contour highlights an insulating $Al_2O_3$ layer, which allows for each electrode to contact only the $WSe_2$ or $WS_2$ respectively. f) Optical image of a $WSe_2$-$MoS_2$ p-n junction device with different electrodes. Panel (c) readapted from Ref. [111] with permission from American Chemical Society (copyright 2014), panel (d) readapted from Ref. [113] with permission from American Chemical Society (copyright 2014), panel (e) readapted from Ref. [70] with permission from Springer Nature and panel (f) readapted from Ref. [114] with permission from The American Association for the Advancement of Science.

#### 3.2.2. Lateral junctions

While in literature one can find many examples of vertical 2D junctions, the same is not true for lateral heterojunctions. In this case a direct mechanical assembly is not possible and one has to rely on bottom-up fabrication methods, as discussed in section 2.2.2. Figure 5e shows a lateral junction device fabricated by Duan *et al.*[70] in which separate contacts are made to $WS_2$ and $WSe_2$. The white dashed rectangle in the image outlines a 50 nm thick $Al_2O_3$ layer deposited onto the $WSe_2$ to insulate the $WS_2$ contact electrodes. Figure 5f shows a different device, fabricated by Li *et al.*,[114] in which different metals respectively Ti and Pd are used to contact the $MoS_2$ and $WSe_2$ regions of the grown flake.

### 3.3. Hybrid and mixed-dimensional

Apart from a purely 2D architecture, many efforts were dedicated to produce hybrid p-n junctions where a 2D material is in contact with a material of lower dimensionality, such as in the case of a molecular crystal or a nanotubes film, or higher dimensionality, like in the case of a bulk semiconductor such as Si. Figure 6a-b show a schematic diagram of a 2D-0D and a 2D-3D p-n junction.

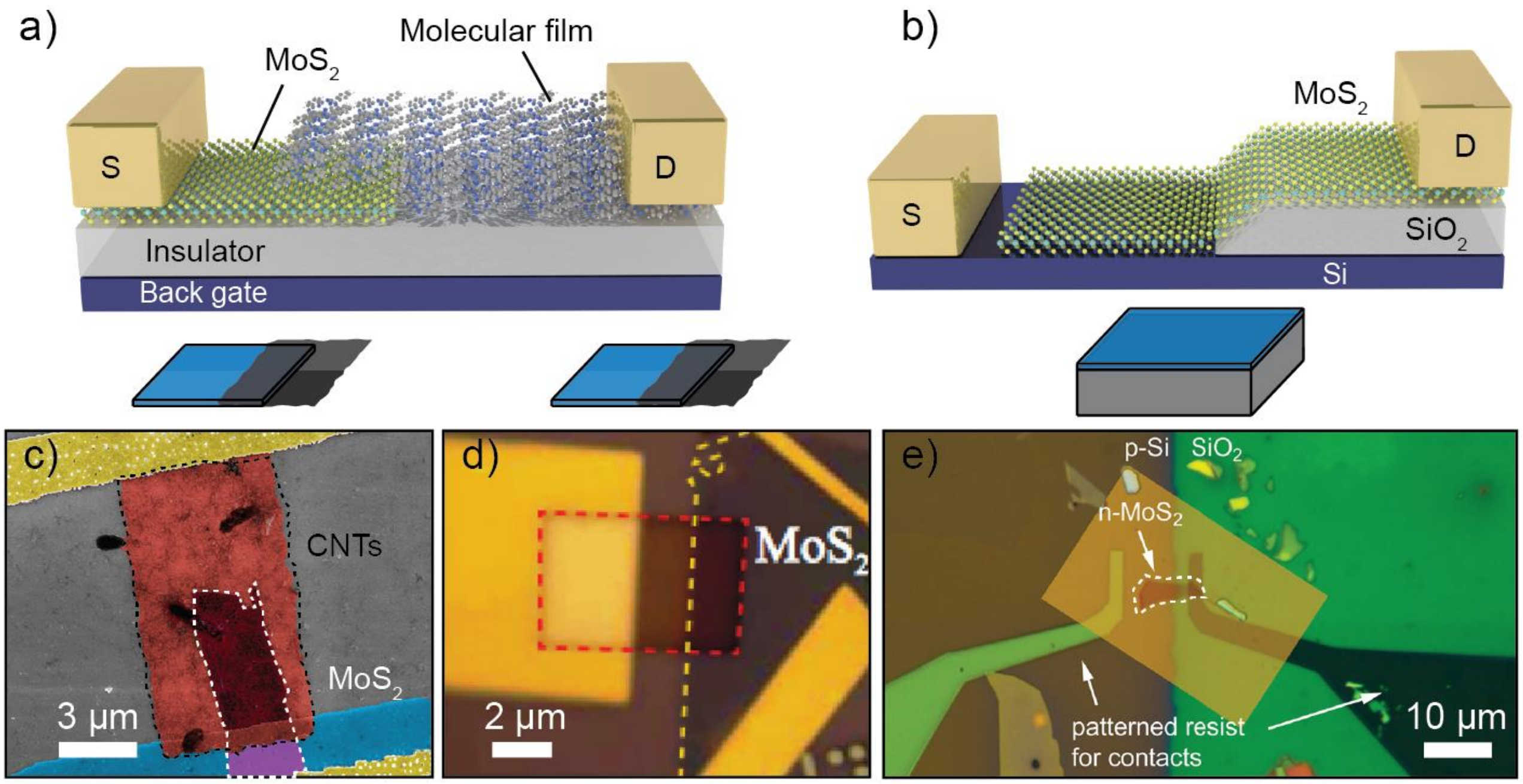

**Figure 6. P-n junctions produced by heterostructures with mixed dimensionality.** a-b) Schematic of mixed-dimensional 2D-0D (a) and 2D-3D (b) heterojunctions. c) False colour SEM image of single wall carbon nanotubes (CNTs) single layer $MoS_2$ p-n junction. d) Optical image of a bilayer $MoS_2$ flake on top of a $SiO_2$/Si substrate to fabricate a 2D-0D heterojunction by depositing a CuPc molecular crystal through a shadow mask. e) Optical image of a 2D-3D device in an intermediate state of fabrication. A monolayer $MoS_2$ is placed across the sidewall of a square window etched into a $SiO_2$ layer exposing the underlying p-doped silicon. Panel (c) readapted from Ref. [115] with permission from National Academy of Sciences, panel (d) readapted from Ref. [116] with permission from Royal Society of Chemistry and panel (e) readapted from Ref. [117] with permission from American Chemical Society.

Molecular organic crystals and nanotube films have surfaces with saturated bonds and are therefore free from dangling bonds. They typically interact via van der Waals forces and this allows the integration of these low-dimensional materials with 2D materials in van der Waals heterostructures.[8, 118] Figure 3c shows a colored scanning electron microscopy (SEM) image of a 1D-2D junction created by transferring a layer of sorted single-walled carbon nanotubes (SWCNTs) onto a single-layer $MoS_2$ flake. A different kind of 1D-2D junctions have been demonstrated by joining a single nanowire and a 2D material. Among the works we find junctions between $MoS_2$, $WSe_2$ and BP and a ZnO nanowire.[119-121] Figure 3d shows a microscope picture of a 0D-2D junction between a bilayer $MoS_2$ flake and a 30 nm thick Cu-phthalocyanine (CuPC) thin film, thermally evaporated through a window opened in a PMMA layer.
Bulk materials such as Si or GaAs, which are largely used in conventional p-n junctions and electronics, can be used in conjunction with 2D materials to create novel p-n 2D-3D junctions. Lopez-Sanchez and coauthors[117] realized a heterojunctions composed of n-type monolayer $MoS_2$ and p-type silicon. Figure 6e shows this $MoS_2$-Si device fabricated by transferring a $MoS_2$ monolayer flake onto a prepatterned highly doped p-type Si substrate covered with $SiO_2$ containing a window through which the underlying Si is exposed. Similar devices based on $MoS_2$ and Si have been thoroughly studied in literature.[122-124] A different geometry was shown by Wang *et al.* [125], who deposited vertically standing $MoS_2$ onto a silicon substrate with a scalable sputtering method. Using different materials, Gehring *et al.*[126] demonstrated a 2D-3D junction composed of a few tens of nm thick black phosphorus flake on top of a highly n-doped GaAs substrate. Multilayer

# 4. Electrical properties

## 4.1. The p-n junction as a rectifier

The first use of a p-n junction that we will discuss is that of a current rectifier. The built-in electric field at the p-n interface allows the flow of charge carriers in one direction (forward bias) while blocking the current in the other direction (backward bias). Many 2D based p-n junctions have been used as rectifiers, showing excellent performances and novel functionalities. One such example is the control of the rectification ratio with a gate field. Deng and coauthors reported gate-tunable rectifying current-voltage characteristics (*IV*s) in a monolayer $MoS_2$-BP p-n junction (see Fig. 5d). Figure 7a shows *IV*s recorded at various gate voltages between -30 V and 50 V. At negative gate voltages they observe a reduction of both the forward current (p-type connected to positive voltage and n-type to negative voltage) and of the reverse current (p-type connected to negative voltage and n-type to positive voltage). Increasing the gate voltage leads to an increase of both the forward and the reverse current in the device. The rectification ratio, defined as the ratio of the forward/reverse current, increases as the back gate voltage decreases as can be seen in Fig. 7b. At a gate voltage of -30 V, a rectification ratio of $10^5$ is obtained. This modulation can be achieved because the band alignment between $MoS_2$ and black phosphorus at the p-n junction interface can be tuned by the gate voltage. Moreover, the gate voltage also modulates the sheet resistance and the contact resistance of the $MoS_2$ and BP respectively.

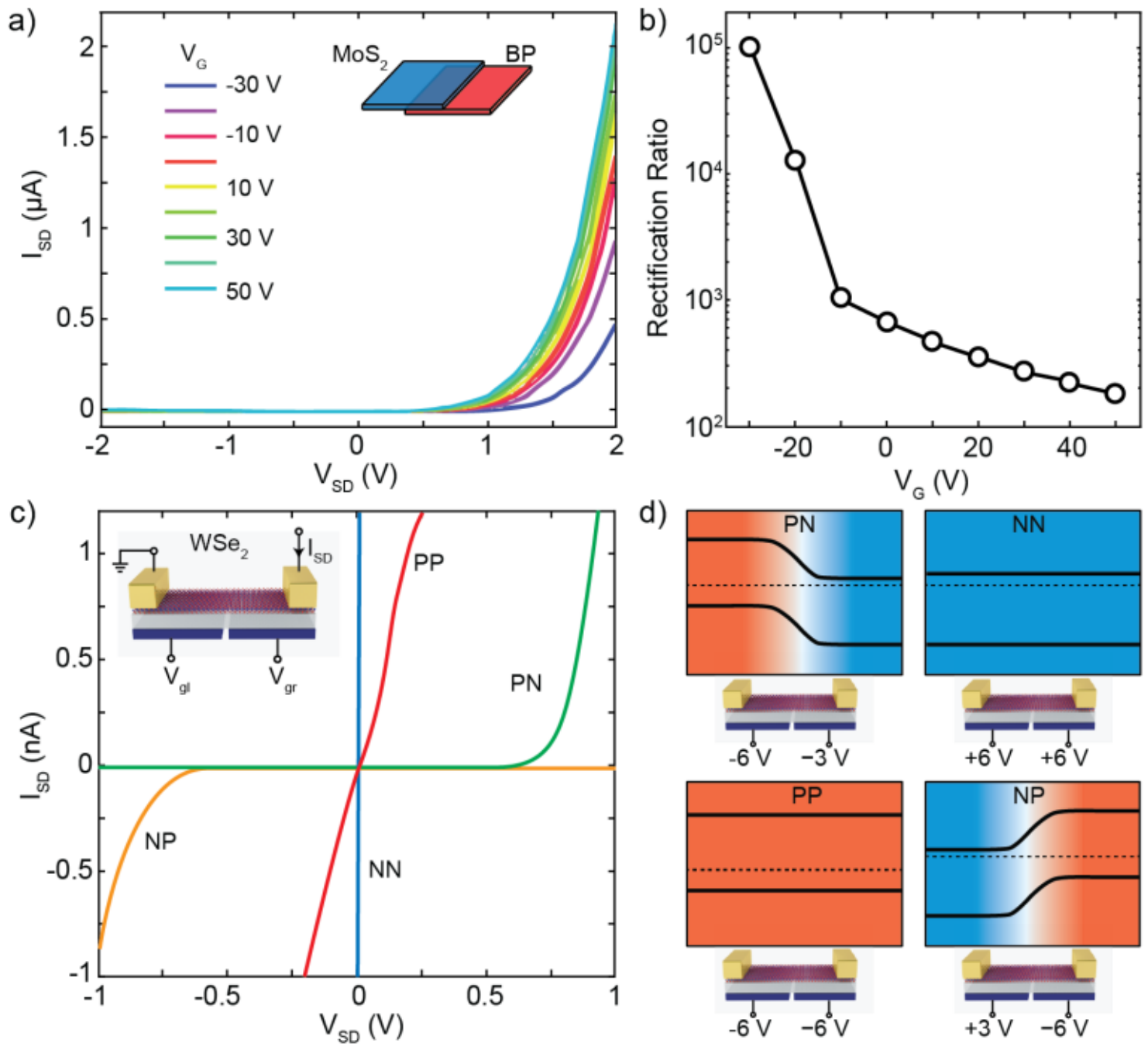

**Figure 7. Atomically thin diodes based on 2D p-n junctions.** a) Gate tunable current-voltage (*IV*) characteristics of a BP-$MoS_2$ vertical heterojunction. b) Rectification ratio at $V_{SD}$ = ±2 V as a function of gate voltage of the device of panel (a). c) *IV* characteristics of a double-gated $WSe_2$ homojunction device in different local gate configurations. d) Schematic diagram of the band structure and the state of the local gates of the device in panel (c) in PP, PN, NP and NN configuration. Panels (a) and (b) readapted from Ref. [113] with permission from American Chemical Society (copyright 2014) and panel (c) readapted from Ref. [79] with permission from American Chemical Society (copyright 2014).

A second kind of devices, that can be called "reconfigurable diode", employs electrostatic doped p-n junctions based on a split gate configuration, which allows inducing locally hole- and electron-doping in different parts of the channel; a new feature introduced by the 2D nature of the devices.[76-78] Figure 7c shows for example the *IV* characteristics of a double-gated monolayer $WSe_2$ homojunction studied by Groenendijk and coauthors.[79] By tuning the Fermi energy both hole- and electron-doping can be readily accessed due to the ambipolarity of $WSe_2$. Two of the four *IV*s of Fig. 7c are linear (NN and PP configuration) while the other two are highly non-linear (PN and NP configuration) displaying rectifying behavior, whose direction can be controlled by the gate bias polarity. The voltages applied to the local gates used to achieve these four configurations are shown in Fig. 7d together with a simplified band diagram of the device.

### 4.2. Physical mechanism of the electrical transport

Although the shape and features of the *IV* characteristics of 2D p-n junctions are similar to those of a conventional p-n junction, the underlying physical mechanism of rectification can be very different. When considering the two model junctions depicted in Fig. 8a one can see that reducing the thickness of the p- and n-type materials modifies the interface between the two materials. In a bulk p-n junction, charge transfer at the interface between the two materials creates a depletion region from which all the free charges are removed. In the case of atomically thin junctions (for example in a vertical junction between monolayer $MoS_2$ and monolayer $WSe_2$) such a depletion region cannot be formed because of the reduced thickness. Simplified diagrams of the band profiles for a multilayer and a monolayer junction, taken along the thickness direction, are shown in Fig. 8a. While the bands of the p- and n-type materials bend in the depletion region in the bulk case (right panels), a sharp discontinuity of the bands is present at the monolayer interface (left panels).
The depletion region in a bulk p-n junction is modified by the application of a voltage as schematically depicted in Fig. 8b. While under forward voltage the size of the depletion region is reduced, under reverse bias its size increases. With increasing forward voltage, the depletion zone eventually becomes thin enough that the built-in electric field cannot counteract charge carrier motion across the p-n junction; this leads to an increase in current. In the case of an atomically thin junction under forward bias the current is governed by tunneling-mediated interlayer recombination between majority carriers at the bottom (top) of the conduction (valence) band of the n-type (p-type) material. This interlayer recombination (Fig. 8c) can be described by two physical mechanisms or a combination of both: Langevin recombination, which is mediated by Coulomb interaction and describes the direct recombination of an electron and a hole, or Shockley–Read–Hall (SRH) recombination, which is mediated by inelastic tunnelling of majority carriers into trap states in the gap.[127-130] These two processes can be present both at the same time in a 2D p-n junction and each of these processes predicts a different dependence of the electron-hole recombination ratio on the majority carriers density. The rectifying *IV*s characteristics can then be explained by an increase of the interlayer recombination rate under forward bias and the photocurrent generated in a 2D junction and its dependence on a gate field can be modeled by the two recombination processes.[112]

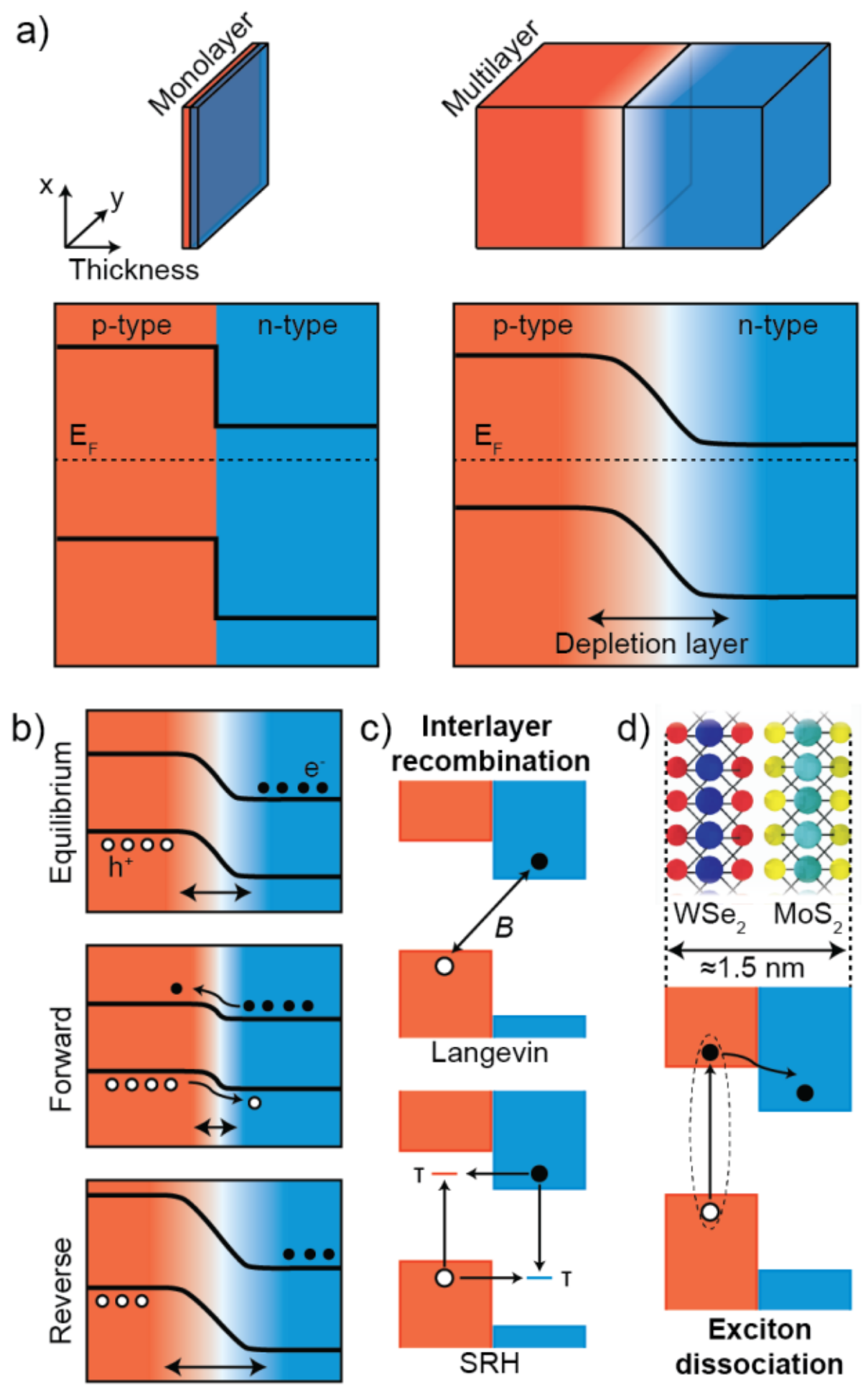

**Figure 8. P-n junctions in bulk and ultra-thin materials.** a) Schematic of the devices (top) and band profiles (bottom) in a multilayer-multilayer p-n junction (right) and monolayer-monolayer (left). In the bottom panels, energy is depicted on the vertical axis and the device thickness on the x-axis. b) Schematic of the bands in a bulk p-n junction under equilibrium ($V_{SD}$=0 V) and under forward and reverse bias. c-d) Schematic diagrams of interlayer recombination (c) in a monolayer-monolayer p-n junction and of (d) the exciton dissociation process.

## 5. Optoelectronic properties

### 5.1. Response to illumination: photodetectors and solar cells

P-n junctions constitute the central building blocks of photodetectors, solar cells and light emitting diodes (LEDs). Their main application is thus in optoelectronics. Typically, there are two operating modes for p-n junctions: photovoltaic mode (PV), in which the p-n junction is not biased, and photoconductive mode (PC), where the p-n junction works under reverse external bias.[131, 132] The PV effect forms the basis for the solar cells. The PC mode is used in photodetection applications and has the advantages of having a lower capacitance that improves the response time and the presence of an external electric field facilitates the separation of electro-hole pairs improving the responsivity.

Some of the most studied vertical p-n junctions are heterostructures made from $WSe_2$ and $MoS_2$.[109, 111, 112] In the work by Furchi *et al.*[111] the authors transferred a mechanically exfoliated monolayer of $WSe_2$ on top of a monolayer $MoS_2$, both on a $SiO_2$/Si substrate, with source and drain electrodes placed in contact with each of the materials. Due to the ambipolar nature of $WSe_2$, this device could be tuned into n-n junction or p-n junction regimes by means of the back gate voltage and, in the p-n configuration, a rectification ratio of ~100 was achieved. By illuminating the device in the p-n configuration with a white light source, a photovoltaic effect was observed, as shown in the *IV*s of Figure 9a measured at increasing optical powers, with an external quantum efficiency (EQE) of ~ 1.5 % and a power conversion efficiency of ~ 0.2 %. In this vertical p-n junction, photons are absorbed in both materials generating electron-hole pairs followed by relaxation of the photogenerated carriers and charge transfer between the layers (see Fig. 8c-d). The carriers have to laterally diffuse to the electrodes and electron-hole recombination may occur reducing the efficiency of the solar cell. Lee *et al.*[112] circumvented this issue by employing graphene electrodes, sandwiching the semiconducting monolayers, and enhancing the responsivity of the devices by a factor of ~5. This increase in responsivity comes mainly from the smaller travel distance that the photoexcited carriers have to travel to reach the electrodes leading to a more efficient carrier extraction in the vertical direction as shown in Fig. 9b. In order to increase the EQE of the devices, solar cells involving materials with different thicknesses were investigated, finding EQEs of 2.4 %, 12 % and 34 % for monolayer, bilayer and multilayer p-n junctions respectively. This improvement is due not only to the enhanced light absorption in the multilayer devices, but also due to the exponential suppression of direct electron tunnelling between the two graphene electrodes. In an optimized (~15 nm thick) $MoS_2$-$WSe_2$ multilayer heterostructure, Wong *et al.*[133] demonstrated internal photocarrier collection efficiencies exceeding 70 % and power conversion efficiencies of up to 3.4 % at 633 nm wavelength.
Furthermore, other combinations of p- and n-doped materials have been explored to realize vertical solar cells, although p-doped 2D materials are less frequent than n-doped materials. Among the naturally p-doped 2D semiconductors we list black phosphorus, franckeite and $MoTe_2$. Deng *et al.* investigated a BP-$MoS_2$ vertical p-n heterojunction working in the visible range of the electromagnetic spectrum with a peak EQE of 0.3 %,[113] a value lower than that of $WSe_2$-$MoS_2$ solar cells. Nevertheless, one of the most attractive features of BP is its low bandgap, allowing light absorption also in the near-infrared part of the electromagnetic spectrum. $MoS_2$-BP junctions have shown EQEs up to ~20 % under illumination with 1.55 µm wavelength light, employing few-layer flakes of both BP and $MoS_2$.[134] More recently, Molina-Mendoza *et al.* investigated the capability of franckeite as near-infrared solar cell in a p-n junction formed by multilayer franckeite-monolayer $MoS_2$ able to generate an electrical power of ~1 pW under infrared illumination with wavelength of 940 nm as shown in Figure 9d.[135] Also, few-layer $MoTe_2$-few-layer $MoS_2$ [136] and monolayer $MoTe_2$-$MoS_2$ [137] p-n junctions have shown promising results in light detection and energy harvesting in the near-infrared. The junction between monolayer $MoTe_2$-$MoS_2$ shows photoresponse at 1550 nm, due to interlayer transitions between the two materials.[137] However, for the individual device made of a pure $MoS_2$ or $MoTe_2$ monolayer, no photoresponse is observed at 1550 nm.
Lateral 2D junctions are also very promising for optoelectronic applications. In the case of heterojunctions, several combinations of materials have already been achieved and studied as energy harvesting devices. CVD grown $WSe_2$-$MoS_2$ was used by Li *et al.* to build a p-n junction which showed an open-circuit voltage of 220 meV under white light illumination.[114] In another work, the authors found for the same system (Figure 9d) a power conversion efficiency of 2.6 % and an increased open-circuit voltage of 390 meV.[138] Devices based on laterally grown $MoSe_2$-$MoS_2$ and $WS_2$-$WSe_2$ heterostructures have also been studied by Duan *et al*, where the authors found that the $WS_2$-$WSe_2$ heterostructures show rectifying behaviour and a photovoltaic effect, yielding an open-circuit voltage of 470 meV and an EQE of ~ 10%.[70]

With respect to homostructures, the most studied materials are $WSe_2$ and BP due to their ambipolar nature that allows for electrostatic doping of different region of the semiconductor channel. Three different groups independently reported a solar cell built by inducing p and n doping in a single-layer $WSe_2$ flake by means of local back-gate electrodes (see Fig. 4b).[76-78] Figure 9e shows *IV*s of this device under illumination in different configurations of the local gates. The *IV*s shows that photogeneration in PV mode occurs only in the NP and PN configurations, with clear short-circuit current and open-circuit voltage. The inset of Fig. 9e shows the photovoltaic power that can be extracted in the PN configuration. Baugher *et al.* and Pospischil *et al.* reported an external quantum efficiency for their solar cells of 0.2 % and 0.5 %, respectively. Alternatively, Buscema *et al.* reported a similar device structure employing BP as the semiconductor channel,[80] which permitted to extend the range of power generation for wavelengths up to 940 nm. On the other hand, the $WSe_2$ devices are limited to the visible region of the electromagnetic spectrum.

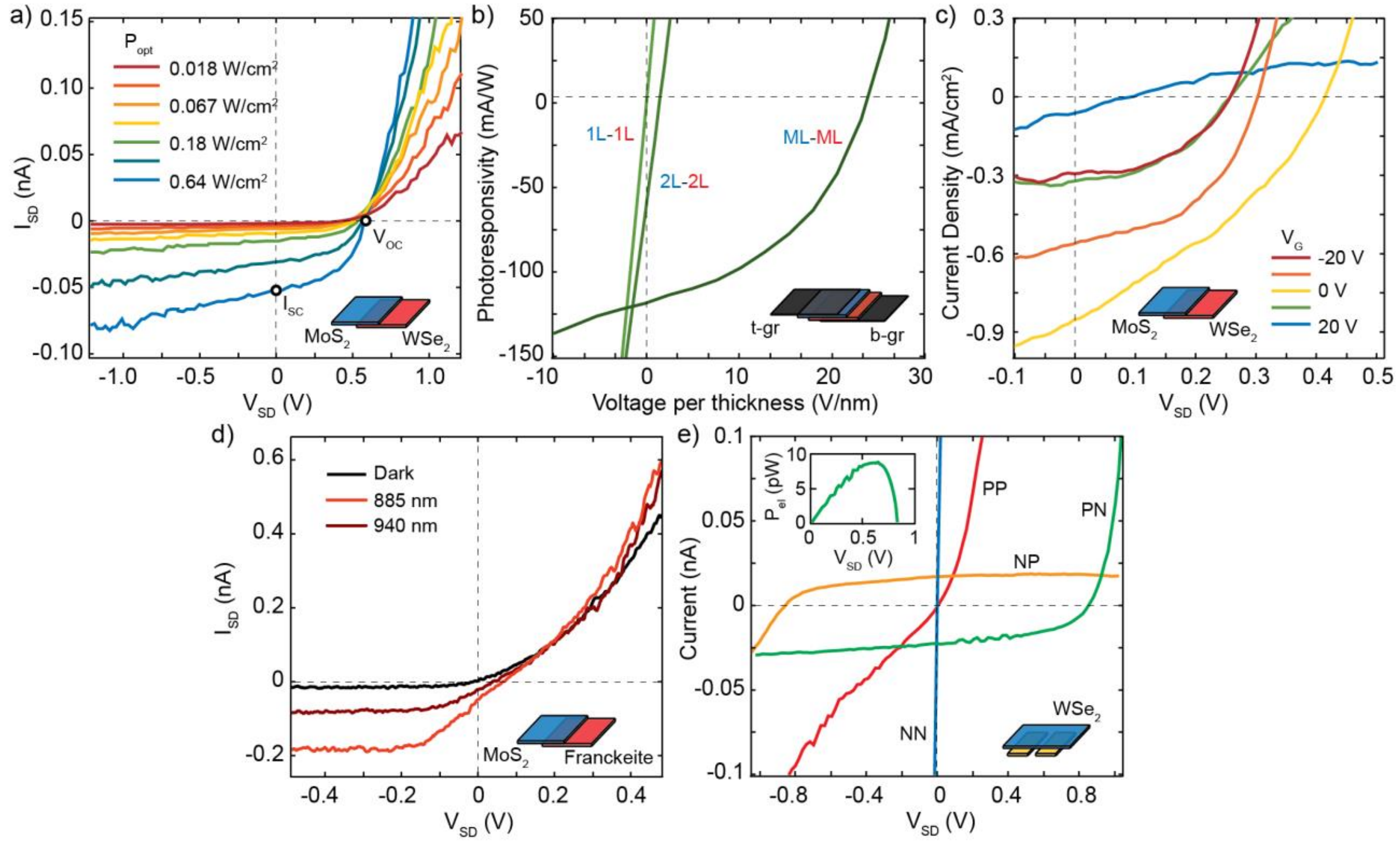

**Figure 9. Optoelectronic properties of 2D p-n junctions.** a) Current-voltage (*IV*) characteristics of a $WSe_2$-$MoS_2$ vertical heterojunction at different illumination powers. b) Photoresponsivity versus voltage traces of $WSe_2$-$MoS_2$ vertical heterojunctions sandwiched between two graphene electrodes for three devices with different thickness of the $WSe_2$ and $MoS_2$ flakes. The x-axis of each trace is normalized by the thickness of the corresponding heterojunction. c) Gate tunable *IV* characteristics of a $WSe_2$-$MoS_2$ vertical heterojunction

kept under illumination. d) *IV* characteristics of a franckeite-$MoS_2$ vertical heterojunction in dark and under infrared illumination. e) *IV* characteristics of a double-gated $WSe_2$ device under optical illumination. The biasing conditions are: PN ($V_{gl}$ = -40 V, $V_{gr}$ = 40 V), NP ($V_{gl}$ = 40 V, $V_{gr}$ = -40 V), NN ($V_{gl}$ = $V_{gr}$ = 40 V), PP ($V_{gl}$ = $V_{gr}$ = -40 V). When operated as a diode (PN and NP), electrical power ($P_{el}$) can be extracted. Inset: *P*el versus voltage in the PN configuration. Panel (a) readapted from Ref. [111] with permission from American Chemical Society, panels (b) and (c) readapted from Ref. [112] with permission from Springer Nature, panel (d) readapted from Ref. [135], panel (e) readapted from Ref. [77] with permission from Springer Nature.

### 5.2. Scanning photocurrent studies

The experiments discussed above were performed under global illumination, that is by illuminating the device with a spot larger than its size. We will now turn to local illumination and specifically to scanning photocurrent (SPC) studies. By scanning a diffraction-limited light spot over a device while measuring the current one can construct SPC maps that yield information on the spatial profile of the bands or the electric fields present in the device.[139-144] This information is crucial to understand the underlying mechanisms ruling the photocurrent generation. The ultrathin nature of 2D p-n junctions allows the light to easily reach the p-n interface both in lateral and in vertical architectures.[64, 78, 145, 146] Figure 10a shows an optical image of a vertical p-n junction composed of elemental doped p-$MoS_2$ and n-$MoS_2$.[64] By recording a SPC map at zero bias the authors studied the local generation of the short circuit current in the device arising from the PV effect. Figure 10b shows such a map recorded with a laser of 532 nm and with highlighted contours of the n- and p-type flakes. A negative photocurrent (blue/red regions) is generated in the region where the two $MoS_2$ flakes overlap as a result of the strong electric field in the depletion region that separates the photocarriers. Compared to global illumination, the SPC map highlights that photocurrent generation can be non-homogeneous across the overlapping region. In Fig. 10b, a hotspot is clearly present, indicating that the interaction between the two flakes is not spatially homogenous, most likely due to the presence of interlayer adsorbates trapped between the layers during the assembly of the device.

SPC studies have also been conducted in lateral devices such as the one depicted in Fig. 10c. Ross *et al.* combined SPC maps with photoluminescence (PL) maps to study an electrostatically gated $WSe_2$ device.[78] The zero bias SPC map, shown in Fig. 10d, has been recorded at 100 K with a 660 nm laser and shows that the photocurrent generation occurs at the interface between the p- and the n-doped regions. Figure 10e presents the corresponding map of the integrated PL intensity, which is homogenous across the whole $WSe_2$ flake, indicating that the luminescence is not quenched by the underlying gates. More revealingly, Fig. 10f shows a color map of the peak photoluminescence photon energy, exhibiting two distinct regions clearly correlated with the n-doped (blue) and p-doped (red) regions of the $WSe_2$ flake. Above the gate held at $V_{gl}$ = +8.0 V the presence of negatively charged $X^-$ trions (two electrons and one hole) dominates, while above the other gate, held at $V_{gr}$ = -8.0 V, the higher-energy positively charged $X^+$ trions (two holes and one electron) are more frequent, implying an excess of holes.

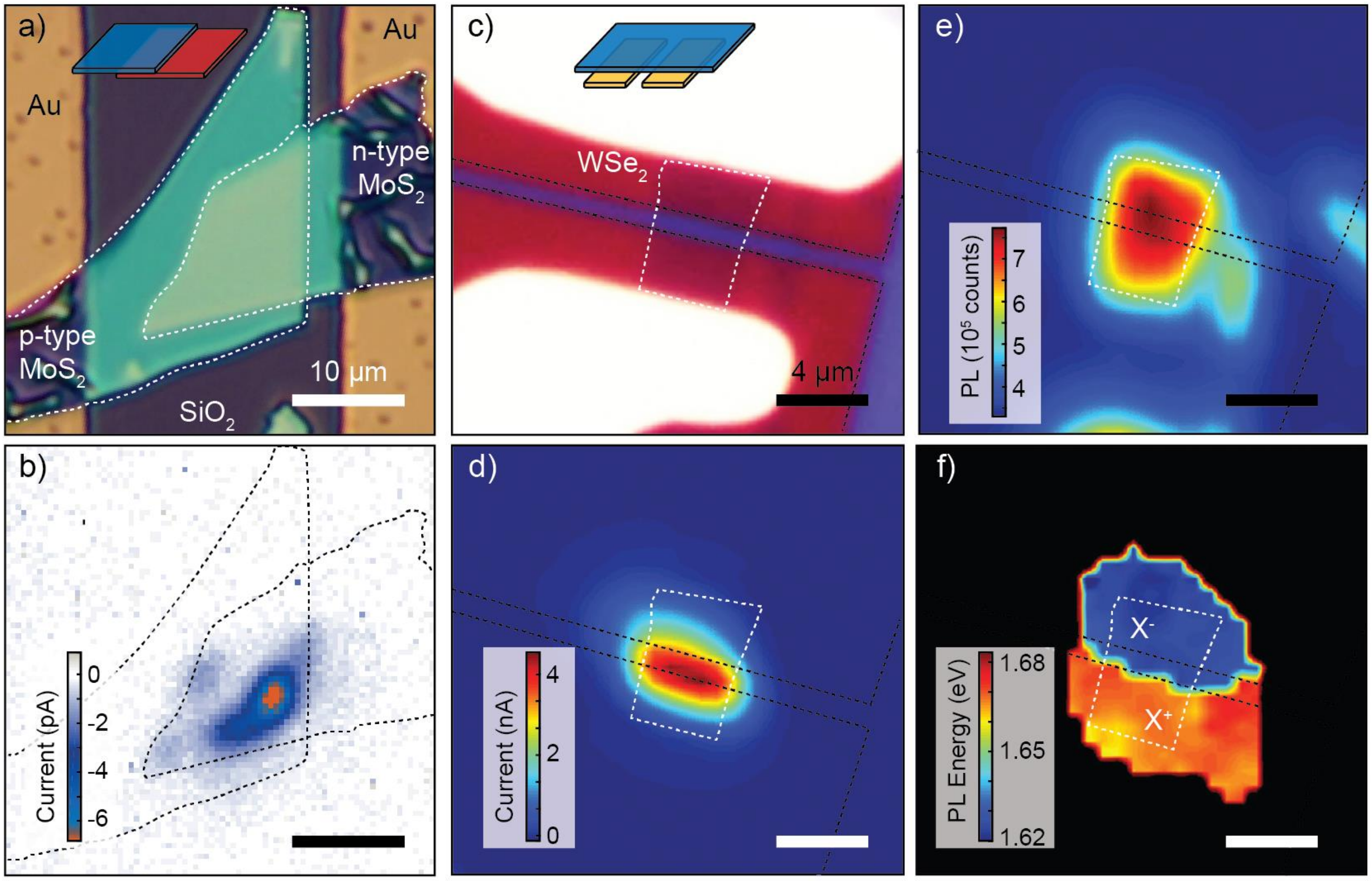

**Figure 10. Scanning photocurrent studies of 2D p-n junctions.** a) Optical image of the device. b) Photocurrent image of the $MoS_2$ p-n junction from panel (a) with zero bias voltage applied (map of the short circuit current). c) Microscope image of a monolayer $WSe_2$ electrostatic p-n junction device. The source and drain contacts are white and the two bottom gates are red. d) Corresponding scanning photocurrent image showing pronounced photocurrent generation localized at the junction. The black dashed lines outline the back gates whereas the white dashed line present the $WSe_2$ flake. e) Integrated photoluminescence map. f) Photoluminescence peak energy map showing p and n regions as a result of the different energies of oppositely charged excitons. $X^-$($X^+$) represents the negatively (positively) charged exciton found in the n (p) region. Panels (a) and (b) readapted from Ref. [64] with permission from John Wiley and Sons and panels (c), (d), (e) and (f) readapted from Ref. [78] with permission from Springer Nature.

## 5.3. Electroluminescence: light emitting diodes

Electroluminescence is the result of radiative recombination of electrons and holes in a material, usually a semiconductor, which release their energy as photons. The most basic requirement for efficient light emission is a direct optical transition in the semiconductor, as in the case of monolayer semiconducting TMDCs. Electrically driven light emission produced by a unipolar current in single layer $MoS_2$ FETs has been reported to take place near the contact electrodes via impact excitation[147] or in a suspended sheet by Joule heating.[148] However, effective emission requires the injection of both electrons and holes, which is typically achieved using a p-n junction.

LEDs based on 2D materials have been realized in different ways, employing different materials and device architectures. For example, the ambipolar nature of single layer $WSe_2$ has been exploited to generate light by employing local gates to electrostatically dope different regions in a semiconductor channel, forming a lateral p-n junction device.[76-78] This kind of device (Figs. 4a and 4d) exhibited electroluminescence efficiencies (ratio between emitted optical power and electrical input power) ranging between 0.1 and 1%, with energy emission peaks at 752 nm (see Fig. 11a, Ref. [77]) and 751 nm (Fig. 11b, Ref. [76]) at room temperature. The difference between the energy of the emission peaks is attributed to different dielectric environments that can influence the exciton binding energy.

Zhang *et al.* have employed mono- and few-layer $WSe_2$ in p-i-n (i, intrinsic) junctions with an electric double-layer transistor (EDLT) architecture,[149] where the doping of the semiconductor channel is modified by tuning the voltage between the drain and source electrodes with respect to the gate electrode, enabling the accumulation of opposite charge carriers close to each of the electrodes. This device configuration circumvents the requirement of using monolayers for light emission due to the breakdown of the inversion symmetry in few-layer flakes under high-gate fields, and it also enables the emission of circularly polarized light as shown in Figure 11c. A similar device structure has been used to induce light emission by means of liquid-gated transistors employing mono- and bilayer $WS_2$,[150] CVD-$MoS_2$ [151] or bulk $ReS_2$.[152]

Concerning electroluminescent heterojunctions and mixed dimensional junctions many examples have been reported in literature. Lopez-Sanchez *et al.* demonstrated the light emission from a monolayer $MoS_2$-highly p-doped Si hybrid junction (see Fig. 6e).[117] Figure 11d shows a gray-scale optical picture of the electroluminescence generated from the device, which exhibits peak emission at ~2 eV. This emission energy is blue-shifted by almost 200 meV from the monolayer $MoS_2$ exciton energy, which is due to the influence of the dielectric environment. The authors also reported that devices with unencapsulated $MoS_2$ show a significant reduction of current and emitted light intensity after a few days in ambient conditions. On the other hand, by encapsulating the $MoS_2$ monolayer with 30 nm thick $HfO_2$ or $Al_2O_3$ the stability of the device can be greatly extended. In a different study Li *et al.* reported EL from a p-i-n junction between multilayer $MoS_2$ and a GaN substrate, using $Al_2O_3$ as insulator.[153]

Furthermore, LEDs based on 2D heterostructures have been reported. Cheng *et al.* measured the light emission from a $WSe_2$/$MoS_2$ vertical heterojunction.[154] In this case, both monolayer-$WSe_2$/few-layer-$MoS_2$ and bilayer-$WSe_2$/few-layer-$MoS_2$ were investigated (Fig. 11e), revealing different features in the emission spectrum related to excitonic peaks A (Fig. 11f) and B (not shown here), hot electron luminescence peaks A' and B' (Fig. 11f) and indirect bandgap emission (not shown here). Table 2 lists some photonic properties of electroluminescent 2D p-n junctions.

| Materials p/n | Peak emission (nm) | Threshold current (nA) | Notes | Reference |
|---|---|---|---|---|
| $WSe_2$ | 752 | 4 | - | [77] |
| $WSe_2$ | 751 | 50 | Estimated electroluminescence efficiency $\eta_{EL} \approx 0.1\%$ | [76] |
| $WSe_2$ | 750 | 5 | - | [78] |
| $WSe_2$ | 740 | 1000 | Emission of circularly polarized light | [149] |
| Si/$MoS_2$ | 620 | 109 | Stable after encapsulation with 30 nm thick oxide | [117] |
| $WSe_2$/$MoS_2$ | 792 | 150000 | - | [154] |
| $WS_2$ | 630 | - | - | [150] |
| $MoS_2$ | 650 | 40000 | Large-area $MoS_2$ monolayer grown by chemical vapor deposition | [151] |

**Table 2**. Photonic properties of on 2D p-n junctions. The threshold current is the lowest current able to produce light emission from the device.

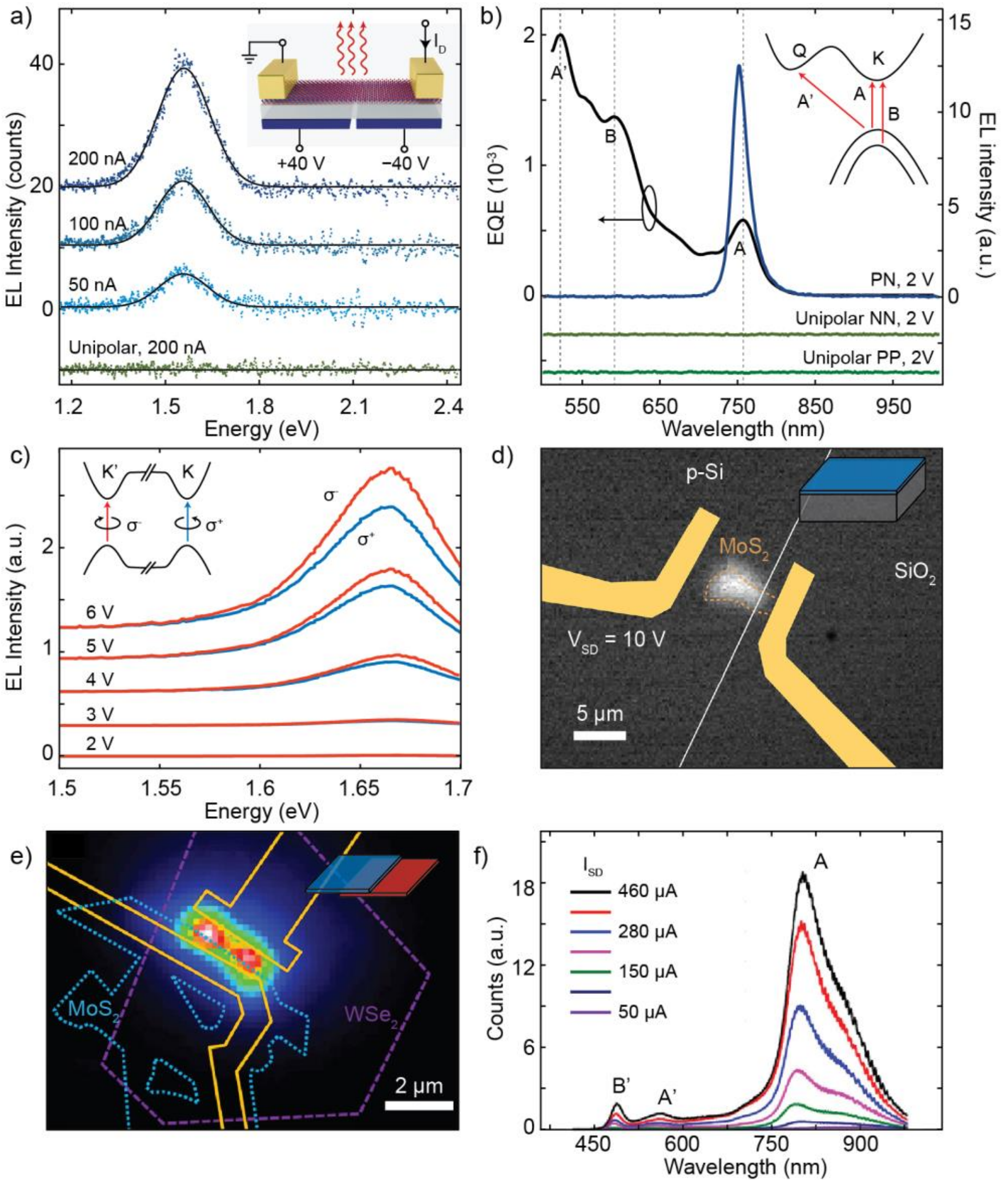

**Figure 11. Light emission from 2D p-n junctions.** a) Electroluminescence (EL) emission spectra of a monolayer $WSe_2$ electrostatic p-n junction recorded for constant currents of 50, 100 and 200 nA. The curves are offset for clarity. The green curve demonstrates that no light emission is obtained under unipolar (n-type) conduction. b) Left axis: external quantum efficiency (EQE) as a function of wavelength at a constant laser power of 2 mW in a monolayer $WSe_2$ device. Peaks in the EQE correspond to exciton transitions A, B and A', as labelled. Right axis: electroluminescence intensity from a monolayer $WSe_2$ electrostatic p-n junction. The traces are offset vertically for clarity. Inset: Diagram of the band structure around the K and Q points, with arrows indicating the lowest-energy exciton transitions for monolayer $WSe_2$. c) Circularly polarized EL spectrum of a $WSe_2$ electric double-layer transistor for different voltages. Inset: Diagram of the band structure around the K and K' points. d) Intensity map showing the electroluminescent emission of a $MoS_2$/Si hybrid

p-n junction. The entire surface of the heterojunction is emitting light. e) False color EL image of a $MoS_2$-$WSe_2$ heterojunction device under an injection current of 100 µA. f) EL spectra of a monolayer $WSe_2/MoS_2$ heterojunction at different injection current. Panel (a) readapted from Ref. [77] with permission from Springer Nature, panel (b) readapted from Ref. [76] with permission from Springer Nature, panel (c) readapted from Ref. [149] with permission from The American Association for the Advancement of Science, panel (d) readapted from Ref. [117] with permission from American Chemical Society, and panels (e) and (f) readapted from Ref. [154] with permission from American Chemical Society.

## 6. Comparison between devices

In this section we show three tables that list the different p-n junctions based on 2D materials found in literature. Table 3 contains the 2D homojunctions, Table 4 the 2D heterojunctions and Table 5 the mixed-dimensional junctions. For each device we report some of the relevant parameters such as the materials composing the device and the thickness. We also list figure of merits for electronic or optoelectronic applications such as the rectification ratio, the responsivity or the open circuit voltage. In the “Materials” column, in the case of heterojunctions and mixed-dimensional junctions the two materials used are listed with the p-type as first and the n-type as second material. In the “Thickness” column the thickness of the junction is given either in nm or in number of layers (indicated by the code nL where n is the number of layers, ML stays for multilayer).

| Homo-junctions | Materials p/n | Thickness (nm) | Rectification ratio | $V_{OC}$ (V) | Reference |
|---|---|---|---|---|---|
| Thickness | $WSe_2$ | 2L/1L | 10 | - | 73 |
| | $MoS_2$ | ML/1L | 1000 | 0 | 74 |
| | $MoSe_2$ | 4/28 | 100000 | 0.24 | 155 |
| | BP | 11L/6L | 600 | 0.21 | 156 |
| Electrostatic | $WSe_2$ | 5.0 | 10000 | 0.83 | 83 |
| | $WSe_2$ | 1L | 50 | 0.03 | 78 |
| | $WSe_2$ | 1L | 500 | 0.85 | 77 |
| | $WSe_2$ | 1L | 100000 | 0.72 | 76 |
| | BP | 6.5 | 11 | 0.05 | 80 |
| | $WSe_2$ | 1L | - | 0.7 | 79 |
| **Chemical** | $MoS_2$ | 3 | 100 | 0.6 | 99 |
| | $MoS_2$ | 60 | 100 | 0.5 | 88 |
| | $MoS_2$ | 7 | 10 | - | 88 |
| | BP | 3 | 100000 | 0.45 | 98 |
| | BP | 10 | 50 | - | 97 |
| **Elemental** | $MoSe_2$ | 6 | 1000000 | 0.35 | 157 |
| | $MoS_2$ | 15 | - | 0.45 | 103 |
| | BP | 8.5 | 5600 | 0.14 | 158 |
| | $MoS_2$ | 10 | 250 | 0.58 | 64 |

**Table 3**. p-n junctions based on 2D homojunctions.

| Hetero-junc-tions | Materials p/n | Thickness (nm) | Rectification ratio | $V_{OC}$ (V) | Reference |
|---|---|---|---|---|---|
| **Vertical** | $WSe_2/MoS_2$ | 1L/1L | 50 | 0.55 | 111 |
| | $WSe_2/MoS_2$ | 1L/1L | 50 | 0.5 | 112 |
| | $WSe_2/MoS_2$ | 2L/13L | 15 | 0.27 | 154 |
| | $BP/MoS_2$ | 11/1L | 100000 | 0.3 | 113 |
| | $WSe_2/MoS_2$ | 5/10 | 50 | 0.28 | 159 |
| | $WSe_2/MoSe_2$ | 1L/1L | 50 | 0.055 | 160 |
| | $WS_2/MoS_2$ | ML/ML | 10000 | 0.25 | 161 |
| | $GaTe/MoS_2$ | 20 | 400000 | 0.22 | 162 |
| | $InAs/WSe_2$ | >20 | 1000000 | - | 163 |
| | $BP/MoS_2$ | 15 | >100 | 0.36 | 164 |
| | $MoTe_2/MoS_2$ | 1L/1L | 10 | 0.15 | 137 |
| | $MoTe_2/MoS_2$ | 4L/4L | 4000 | 0.3 | 136 |
| | $WSe_2/MoSe_2$ | 3L/3L | 10000 | 0.46 | 165 |
| | $WSe_2$/BP | 12/20 | 2500 | 0.29 | 166 |
| | franckeite/$MoS_2$ | 25/1.4 | 400 | 0.08 | 135 |
| | $ReSe_2$/$MoS_2$ | 60/7 | 60000 | 0.1 | 167 |
| | $ReS_2$/$ReSe_2$ | 64/48 | 3150 | 0.18 | 168 |
| | GaSe/InSe | 19/13 | 200000 | - | 169 |
| | GeSe/$MoS_2$ | - | >100 | - | 170 |
| | $WSe_2$/$MoS_2$ | 9 | 1000 | - | 171 |
| | $MoTe_2$/$SnSe_2$ | 31 | 1000 | - | 171 |
| | SnSe/$MoS_2$ | 28/7 | 100000 | - | 172 |
| | SnS/$WS_2$ | 200/0.7 | 15 | - | 173 |
| | $WSe_2$/$SnS_2$ | 1L/1L | 10000000 | 0.03 | 174 |
| | CuO/$MoS_2$ | 150/1L | 10 | - | 175 |
| Lateral | $WS_2$/$MoS_2$ | 1L | 100 | 0.12 | 65 |
| | $WSe_2$/$WS_2$ | ML | - | 0.47 | 70 |
| | $In_2Se_3$/$CuInSe_2$ | 14 | 10 | 0.03 | 72 |
| | $WSe_2$/$MoS_2$ | 1L | 10 | 0.22 | 114 |

**Table 4**. p-n junctions based on 2D heterojunctions.

| Hetero-junc-tions | Materials p/n | Thickness (nm) | Rectification ratio | $V_{OC}$ (V) | Reference |
|---|---|---|---|---|---|
| 2D-0D, 2D-1D | SWCNT/$MOS_2$ | -/1L | 10000 | - | 115 |
| | Rubrene/$MoS_2$ | 300/5 | 100000 | - | 176 |
| | Pentacene/$MoS_2$ | 40/2L | 5 | 0.3 | 177 |
| | $C_8$-BTBT/$MoS_2$ | 5 | 100000 | 0.5 | 178 |
| | CuPC/$MoS_2$ | 20/1L | 10000 | 0.6 | 116 |
| **2D-3D** | Si/$MoS_2$ | 1L | - | 0.41 | [122] |
| | Ge/$MoS_2$ | 2L | 2 | - | [179] |
| | Si/$MoS_2$ | 1L | 100 | 0.58 | [117] |
| | BP/GaAs | 15 | 120 | 0.55 | [126] |
| | $WS_2$/GaN | 400 | 1000 | - | [180] |
| | Si/$MoS_2$ | 12.5 | 20 | - | [181] |
| | Si/$Bi_2Se_3$ | - | 50 | 0.24 | [182] |
| | LSMO/$MoS_2$ | 1L | 1000 | 0.4 | [183] |

**Table 5**. p-n junctions based on mixed dimensional heterojunctions.

## 7. Future perspective and challenges

Despite the large number of experiments discussed in this Review, many challenges still need to be overcome for the integration of 2D p-n junctions in mass-produced electronic components. The two most important challenges are the large-scale fabrication of 2D p-n junctions and the environmental degradation of 2D materials.

### 7.1. Large-scale fabrication and environmental degradation

The first main challenge is related to the large-scale production of tailored van der Waals heterostructures with well-controlled interfaces. Even if the deterministic placement methods are very successful for laboratory-scale experiments, they are not suited for commercial applications. Growing methods like the CVD growth described in Section 2.2.2 have already proven to be capable of growing high-quality 2D materials including lateral and vertical heterostructures at a laboratory-scale. Van der Waals epitaxial methods hold even more promises to the synthesis of high-quality 2D heterostructures. Up-scaling of these growth methods is possible and in the years to come the realization of higher quality devices can be expected.[184] A second, promising strategy to up-scale the production of 2D p-n junctions is to combine the growth of single 2D materials (such as $MoS_2$) with various doping techniques (mostly chemical or electrostatic).

A second challenge is the environmental degradation of many of the known 2D materials. For example, when exposed to air, black phosphorous in its ultrathin form tends to uptake moisture which degrades the electronic properties of the material.[37-40] In the case of BP the most accepted mechanism for the degradation involves the reaction of the material with oxygen which changes the material properties.[185, 186] One way prevent this degradation is by encapsulating the air-sensitive material between two flakes of h-BN under oxygen- and moisture-free conditions.[187-189] One active area of 2D materials research is therefore dedicated to up-scale the encapsulation methods. A different approach that is currently pursued is the active search for new 2D materials, which do not present degradation problems, that could come either from synthesis (for example $TiS_3$) or from natural sources (*e.g.* franckeite).[46, 135, 190, 191] This active search already helped to increase the number of available 2D materials from just a handful to more than twenty in less than ten years.

### 7.2. Future perspectives

Apart from conventional optoelectronic applications, 2D p-n junctions hold still many unexplored applications and fundamental questions. For example, thermoelectric applications of 2D p-n junctions have not been thoroughly investigated yet. The traditional Peltier device, a component largely used in electronics for cooling (and less commonly for heating), is based on a p-type and a n-type semiconductor thermally connected in parallel and electrically in series. Van der Waals heterostructures could be used to fabricate atomically thin cooling (or heating) elements combined with the other attractive properties of 2D materials, such as the high transparence or the flexibility. Another interesting route is the study of novel p-n junction geometries (like for example circular p-n junctions recently demonstrated in graphene[192, 193]) or new devices based on 2D p-n junctions, such as memories or logic gates.[83, 194]

The real possibilities still hidden in 2D materials are probably many more than the ones discussed and 2D p-n junctions hold many promises for commercial applications. These 2D junctions are especially interesting building blocks for flexible and transparent electronic components, such as solar cells or light emitting diodes.[195, 196] Another application that can benefit from the ultrathin nature of 2D p-n junctions is in light sensing and harvesting applications or for nanophotonics.[14, 197] As we discussed in section 5, 2D p-n junctions can be used as photodetectors and many materials combinations are available that can be used to design devices sensible to wavelengths ranging from the infrared to the ultraviolet have already been demonstrated.[134, 135, 137]

## 8. Conclusions

In recent years we have witnessed the production of novel p-n junctions that take advantage of the ultrathin nature of 2D materials. Both bottom-up and top-down production methods have proven capable of creating p-n junctions of high quality with exquisite optoelectronic properties. In this Review we have revised the recent progress on 2D p-n junctions, discussing the most used materials and fabrication methods and examples of 2D p-n junctions from literature belonging to eight different junction architectures. With these architectures different applications have been realized and we have discussed experiments that use 2D p-n junctions respectively as rectifiers, as photodetectors/solar cells and as light emitting diodes. Finally a comparison between the important figures of merit of the various devices from literature is made. By no means, this field is closed. 2D materials continue to offer many opportunities to fabricate novel p-n junctions with outstanding properties that open up exciting scientific routes both in terms of fundamental questions and in term of applications.

## ACKNOWLEDGEMENTS

AC-G acknowledges funding from the European Commission under the Graphene Flagship, contract CNECTICT-604391. RF acknowledges support from the Netherlands Organization for Scientific Research (NWO) through the research program Rubicon with project number 680-50-1515. This project further received funding from the European Union's Horizon 2020 research and innovation program under grant agreement No. 696656 (Graphene Flagship).

## COMPETING INTERESTS

The authors declare no competing financial interests.

## REFERENCES

1. M. Riordan and L. Hoddeson, *IEEE spectrum*, 1997, **34**, 46-51.
2. Q. H. Wang, K. Kalantar-Zadeh, A. Kis, J. N. Coleman and M. S. Strano, *Nature Nanotechnology*, 2012, **7**, 699-712.
3. S. J. Kim, K. Choi, B. Lee, Y. Kim and B. H. Hong, *Annual Review of Materials Research*, 2015, **45**, 63-84.
4. H. Schmidt, F. Giustiniano and G. Eda, *Chemical Society Reviews*, 2015, **44**, 7715-7736.
5. S. Das, J. A. Robinson, M. Dubey, H. Terrones and M. Terrones, *Annual Review of Materials Research*, 2015, **45**, 1-27.
6. K. Novoselov, A. Mishchenko, A. Carvalho and A. C. Neto, *Science*, 2016, **353**, aac9439.
7. Y. Liu, N. O. Weiss, X. Duan, H.-C. Cheng, Y. Huang and X. Duan, *Nature Reviews Materials*, 2016, **1**, 16042.
8. D. Jariwala, T. J. Marks and M. C. Hersam, *Nature materials*, 2016, **16**, 170-181.
9. K. S. Novoselov, A. K. Geim, S. V. Morozov, D. Jiang, Y. Zhang, S. V. Dubonos, I. V. Grigorieva and A. A. Firsov, *Science*, 2004, **306**, 666-669.
10. V. Nicolosi, M. Chhowalla, M. G. Kanatzidis, M. S. Strano and J. N. Coleman, *Science*, 2013, **340**, 1226419.
11. K. Novoselov, D. Jiang, F. Schedin, T. Booth, V. Khotkevich, S. Morozov and A. Geim, *Proceedings of the National Academy of Sciences of the United States of America*, 2005, **102**, 10451-10453.
12. K. Novoselov and A. C. Neto, *Physica Scripta*, 2012, **2012**, 014006.
13. P. Miró, M. Audiffred and T. Heine, *Chemical Society Reviews*, 2014, **43**, 6537-6554.
14. A. Castellanos-Gomez, *Nature Photonics*, 2016, **10**, 202-204.
15. R. Roldan, L. Chirolli, E. Prada, J. A. Silva-Guillen, P. San-Jose and F. Guinea, *Chemical Society Reviews*, 2017, **46**, 4387-4399.
16. B. Partoens and F. Peeters, *Physical Review B*, 2006, **74**, 075404.
17. M. Y. Han, B. Özyilmaz, Y. Zhang and P. Kim, *Physical review letters*, 2007, **98**, 206805.
18. K. F. Mak, C. H. Lui, J. Shan and T. F. Heinz, *Physical review letters*, 2009, **102**, 256405.
19. D. C. Elias, R. R. Nair, T. Mohiuddin, S. Morozov, P. Blake, M. Halsall, A. Ferrari, D. Boukhvalov, M. Katsnelson and A. Geim, *Science*, 2009, **323**, 610-613.
20. F. Xia, D. B. Farmer, Y.-m. Lin and P. Avouris, *Nano letters*, 2010, **10**, 715-718.
21. R. Balog, B. Jørgensen, L. Nilsson, M. Andersen, E. Rienks, M. Bianchi, M. Fanetti, E. Lægsgaard, A. Baraldi and S. Lizzit, *Nature materials*, 2010, **9**, 315-319.
22. Y. Lin and J. W. Connell, *Nanoscale*, 2012, **4**, 6908-6939.
23. M. Chhowalla, H. S. Shin, G. Eda, L.-J. Li, K. P. Loh and H. Zhang, *Nature chemistry*, 2013, **5**, 263.
24. D. Jariwala, V. K. Sangwan, L. J. Lauhon, T. J. Marks and M. C. Hersam, *ACS nano*, 2014, **8**, 1102-1120.
25. X. Duan, C. Wang, A. Pan, R. Yu and X. Duan, *Chemical Society Reviews*, 2015, **44**, 8859-8876.
26. O. V. Yazyev and A. Kis, *Materials Today*, 2015, **18**, 20-30.
27. W. Choi, N. Choudhary, G. H. Han, J. Park, D. Akinwande and Y. H. Lee, *Materials Today*, 2017.
28. S. Manzeli, D. Ovchinnikov, D. Pasquier, O. V. Yazyev and A. Kis, *Nature Reviews Materials*, 2017, **2**, 17033.
29. H. Li, J. Wu, Z. Yin and H. Zhang, *Accounts of chemical research*, 2014, **47**, 1067-1075.
30. K. F. Mak, C. Lee, J. Hone, J. Shan and T. F. Heinz, *Physical Review Letters*, 2010, **105**, 136805.
31. A. Splendiani, L. Sun, Y. Zhang, T. Li, J. Kim, C.-Y. Chim, G. Galli and F. Wang, *Nano Letters*, 2010, **10**, 1271-1275.
32. B. Radisavljevic, A. Radenovic, J. Brivio, V. Giacometti and A. Kis, *Nature Nanotechnology*, 2011, **6**, 147-150.
33. F. Xia, H. Wang and Y. Jia, *Nature communications*, 2014, **5**, 4458.
34. J. Yang, R. Xu, J. Pei, Y. W. Myint, F. Wang, Z. Wang, S. Zhang, Z. Yu and Y. Lu, *Light: Science & Applications*, 2015, **4**, e312.
35. R. Gusmao, Z. Sofer and M. Pumera, *Angewandte Chemie*, 2017.
36. A. Castellanos-Gomez, L. Vicarelli, E. Prada, J. O. Island, K. Narasimha-Acharya, S. I. Blanter, D. J. Groenendijk, M. Buscema, G. A. Steele and J. Alvarez, *2D Materials*, 2014, **1**, 025001.
37. J. O. Island, G. A. Steele, H. S. van der Zant and A. Castellanos-Gomez, *2D Materials*, 2015, **2**, 011002.
38. J. D. Wood, S. A. Wells, D. Jariwala, K.-S. Chen, E. Cho, V. K. Sangwan, X. Liu, L. J. Lauhon, T. J. Marks and M. C. Hersam, *Nano letters*, 2014, **14**, 6964-6970.
39. J.-S. Kim, Y. Liu, W. Zhu, S. Kim, D. Wu, L. Tao, A. Dodabalapur, K. Lai and D. Akinwande, *Scientific reports*, 2015, **5**.
40. A. Favron, E. Gaufrès, F. Fossard, A.-L. Phaneuf-L'Heureux, N. Y. Tang, P. L. Lévesque, A. Loiseau, R. Leonelli, S. Francoeur and R. Martel, *Nature materials*, 2015, **14**, 826-832.
41. A. Molle, J. Goldberger, M. Houssa, Y. Xu, S.-C. Zhang and D. Akinwande, *Nature materials*, 2017.
42. S. Balendhran, S. Walia, H. Nili, S. Sriram and M. Bhaskaran, *small*, 2015, **11**, 640-652.

43. L.-D. Zhao, S.-H. Lo, Y. Zhang, H. Sun, G. Tan, C. Uher, C. Wolverton, V. P. Dravid and M. G. Kanatzidis, *Nature*, 2014, **508**, 373-377.
44. C. W. Li, J. Hong, A. F. May, D. Bansal, S. Chi, T. Hong, G. Ehlers and O. Delaire, *Nature Physics*, 2015, **11**, 1063-1069.
45. S. Srivastava and B. Avasthi, *Journal of materials science*, 1992, **27**, 3693-3705.
46. J. O. Island, A. J. Molina-Mendoza, M. Barawi, R. Biele, E. Flores, J. M. Clamagirand, J. R. Ares, C. Sánchez, H. S. van der Zant and R. D'Agosta, *2D Materials*, 2017, **4**, 022003.
47. B. Huang, G. Clark, E. Navarro-Moratalla, D. R. Klein, R. Cheng, K. L. Seyler, D. Zhong, E. Schmidgall, M. A. McGuire, D. H. Cobden, W. Yao, D. Xiao, P. Jarillo-Herrero and X. Xu, *Nature*, 2017, **546**, 270-273.
48. M. A. McGuire, *Crystals*, 2017, **7**, 121.
49. M. Osada and T. Sasaki, *Journal of Materials Chemistry*, 2009, **19**, 2503-2511.
50. R. Ma and T. Sasaki, *Accounts of chemical research*, 2014, **48**, 136-143.
51. B. Anasori, M. R. Lukatskaya and Y. Gogotsi, *Nature Reviews Materials*, 2017, **2**, 16098.
52. J. Kang, S. Tongay, J. Zhou, J. Li and J. Wu, *Applied Physics Letters*, 2013, **102**, 012111.
53. Y. Liu, P. Stradins and S.-H. Wei, *Science advances*, 2016, **2**, e1600069.
54. V. Tran, R. Soklaski, Y. Liang and L. Yang, *Physical Review B*, 2014, **89**, 235319.
55. Y. Cai, G. Zhang and Y.-W. Zhang, *Scientific reports*, 2014, **4**, 6677.
56. P. Blake, E. Hill, A. Castro Neto, K. Novoselov, D. Jiang, R. Yang, T. Booth and A. Geim, *Applied Physics Letters*, 2007, **91**, 063124.
57. Z. Ni, H. Wang, J. Kasim, H. Fan, T. Yu, Y. Wu, Y. Feng and Z. Shen, *Nano letters*, 2007, **7**, 2758-2763.
58. S. Roddaro, P. Pingue, V. Piazza, V. Pellegrini and F. Beltram, *Nano letters*, 2007, **7**, 2707-2710.
59. R. Frisenda, E. Navarro-Moratalla, P. Gant, D. P. De Lara, P. Jarillo-Herrero, R. V. Gorbachev and A. Castellanos-Gomez, *Chemical Society Reviews*, 2018, **47**, 53-68.
60. C. R. Dean, A. F. Young, I. Meric, C. Lee, L. Wang, S. Sorgenfrei, K. Watanabe, T. Taniguchi, P. Kim and K. L. Shepard, *Nature nanotechnology*, 2010, **5**, 722-726.
61. P. Zomer, S. Dash, N. Tombros and B. Van Wees, *Applied Physics Letters*, 2011, **99**, 232104.
62. A. Castellanos-Gomez, M. Buscema, R. Molenaar, V. Singh, L. Janssen, H. S. van der Zant and G. A. Steele, *2D Materials*, 2014, **1**, 011002.
63. L. Wang, I. Meric, P. Huang, Q. Gao, Y. Gao, H. Tran, T. Taniguchi, K. Watanabe, L. Campos and D. Muller, *Science*, 2013, **342**, 614-617.
64. C. Reuter, R. Frisenda, D. Y. Lin, T. S. Ko, D. Perez de Lara and A. Castellanos-Gomez, *Small Methods*, 2017, **1**, 1700119.
65. Y. Gong, J. Lin, X. Wang, G. Shi, S. Lei, Z. Lin, X. Zou, G. Ye, R. Vajtai and B. I. Yakobson, *Nature materials*, 2014, **13**, 1135-1142.
66. J. Yu, J. Li, W. Zhang and H. Chang, *Chemical Science*, 2015, **6**, 6705-6716.
67. S. L. Wong, H. Liu and D. Chi, *Progress in Crystal Growth and Characterization of Materials*, 2016, **62**, 9-28.
68. R. Dong and I. Kuljanishvili, *Journal of Vacuum Science & Technology B, Nanotechnology and Microelectronics: Materials, Processing, Measurement, and Phenomena*, 2017, **35**, 030803.
69. Y. Cui, B. Li, J. Li and Z. Wei, *Science China Physics, Mechanics & Astronomy*, 2018, **61**, 016801.
70. X. Duan, C. Wang, J. C. Shaw, R. Cheng, Y. Chen, H. Li, X. Wu, Y. Tang, Q. Zhang and A. Pan, *Nature nanotechnology*, 2014, **9**, 1024-1030.
71. C. Huang, S. Wu, A. M. Sanchez, J. J. Peters, R. Beanland, J. S. Ross, P. Rivera, W. Yao, D. H. Cobden and X. Xu, *Nature materials*, 2014, **13**, 1096-1101.
72. Z. Zheng, J. Yao and G. Yang, *ACS Applied Materials & Interfaces*, 2017, **9**, 7288-7296.
73. Z.-Q. Xu, Y. Zhang, Z. Wang, Y. Shen, W. Huang, X. Xia, W. Yu, Y. Xue, L. Sun and C. Zheng, *2D Materials*, 2016, **3**, 041001.
74. M. Sun, D. Xie, Y. Sun, W. Li, C. Teng and J. Xu, *Scientific Reports*, 2017, **7**, 4505.
75. R. Bratschitsch, *Nature nanotechnology*, 2014, **9**, 247-248.
76. B. W. Baugher, H. O. Churchill, Y. Yang and P. Jarillo-Herrero, *Nature nanotechnology*, 2014, **9**, 262-267.
77. A. Pospischil, M. M. Furchi and T. Mueller, *Nature nanotechnology*, 2014, **9**, 257-261.
78. J. S. Ross, P. Klement, A. M. Jones, N. J. Ghimire, J. Yan, D. Mandrus, T. Taniguchi, K. Watanabe, K. Kitamura and W. Yao, *Nature nanotechnology*, 2014, **9**, 268-272.
79. D. J. Groenendijk, M. Buscema, G. A. Steele, S. Michaelis de Vasconcellos, R. Bratschitsch, H. S. van der Zant and A. Castellanos-Gomez, *Nano letters*, 2014, **14**, 5846-5852.
80. M. Buscema, D. J. Groenendijk, G. A. Steele, H. S. J. van der Zant and A. Castellanos-Gomez, *Nature Communications*, 2014, **5**, 4651.
81. N. M. Gabor, J. C. Song, Q. Ma, N. L. Nair, T. Taychatanapat, K. Watanabe, T. Taniguchi, L. S. Levitov and P. Jarillo-Herrero, *Science*, 2011, **334**, 648-652.
82. H.-Y. Chiu, V. Perebeinos, Y.-M. Lin and P. Avouris, *Nano letters*, 2010, **10**, 4634-4639.

83. D. Li, M. Chen, Z. Sun, P. Yu, Z. Liu, P. M. Ajayan and Z. Zhang, *Nature Nanotechnology*, 2017.
84. J. Ye, S. Inoue, K. Kobayashi, Y. Kasahara, H. Yuan, H. Shimotani and Y. Iwasa, *Nature materials*, 2010, **9**.
85. Y. Zhang, J. Ye, Y. Matsuhashi and Y. Iwasa, *Nano letters*, 2012, **12**, 1136-1140.
86. Y. Zhang, J. Ye, Y. Yomogida, T. Takenobu and Y. Iwasa, *Nano letters*, 2013, **13**, 3023-3028.
87. A. A. El Yumin, J. Yang, Q. Chen, O. Zheliuk and J. Ye, *physica status solidi (b)*, 2017, **254**, 1700180.
88. M. S. Choi, D. Qu, D. Lee, X. Liu, K. Watanabe, T. Taniguchi and W. J. Yoo, *ACS nano*, 2014, **8**, 9332-9340.
89. S. Mouri, Y. Miyauchi and K. Matsuda, *Nano letters*, 2013, **13**, 5944-5948.
90. J. D. Lin, C. Han, F. Wang, R. Wang, D. Xiang, S. Qin, X. A. Zhang, L. Wang, H. Zhang, A. T. Wee and W. Chen, *ACS Nano*, 2014, **8**, 5323-5329.
91. C. R. Ryder, J. D. Wood, S. A. Wells, Y. Yang, D. Jariwala, T. J. Marks, G. C. Schatz and M. C. Hersam, *Nature chemistry*, 2016, **8**, 597-602.
92. G. Abellán, V. Lloret, U. Mundloch, M. Marcia, C. Neiss, A. Görling, M. Varela, F. Hauke and A. Hirsch, *Angewandte Chemie International Edition*, 2016, **55**, 14557-14562.
93. A. J. Molina-Mendoza, L. Vaquero-Garzon, S. Leret, L. de Juan-Fernández, E. M. Pérez and A. Castellanos-Gomez, *Chemical Communications*, 2016, **52**, 14365-14368.
94. X. He, W. Chow, F. Liu, B. Tay and Z. Liu, *Small*, 2017, **13**, 1602558.
95. E. C. Peters, E. J. Lee, M. Burghard and K. Kern, *Applied Physics Letters*, 2010, **97**, 193102.
96. D. B. Farmer, Y.-M. Lin, A. Afzali-Ardakani and P. Avouris, *Applied Physics Letters*, 2009, **94**, 213106.
97. X. Yu, S. Zhang, H. Zeng and Q. J. Wang, *Nano Energy*, 2016, **25**, 34-41.
98. G. Wang, L. Bao, T. Pei, R. Ma, Y.-Y. Zhang, L. Sun, G. Zhang, H. Yang, J. Li and C. Gu, *Nano letters*, 2016, **16**, 6870-6878.
99. H.-M. Li, D. Lee, D. Qu, X. Liu, J. Ryu, A. Seabaugh and W. J. Yoo, *Nature communications*, 2015, **6**.
100. J. Legma, G. Vacquier, H. Traore and A. Casalot, *Materials Science and Engineering: B*, 1991, **8**, 167-174.
101. J. Suh, T. E. Park, D. Y. Lin, D. Fu, J. Park, H. J. Jung, Y. Chen, C. Ko, C. Jang, Y. Sun, R. Sinclair, J. Chang, S. Tongay and J. Wu, *Nano Lett*, 2014, **14**, 6976-6982.
102. Y. Jin, D. H. Keum, S. J. An, J. Kim, H. S. Lee and Y. H. Lee, *Adv Mater*, 2015, **27**, 5534-5540.
103. S. A. Svatek, E. Antolín, D.-Y. Lin, R. Frisenda, C. Reuter, A. J. Molina-Mendoza, M. Muñoz, N. Agrait, T.-S. Ko, D. P. de Lara and A. Castellanos-Gomez, *Journal of Materials Chemistry C*, 2017, **5**, 854-861.
104. A. Nipane, D. Karmakar, N. Kaushik, S. Karande and S. Lodha, *ACS nano*, 2016, **10**, 2128-2137.
105. E. Kim, C. Ko, K. Kim, Y. Chen, J. Suh, S. G. Ryu, K. Wu, X. Meng, A. Suslu and S. Tongay, *Advanced Materials*, 2016, **28**, 341-346.
106. A. K. Geim and I. V. Grigorieva, *Nature*, 2013, **499**, 419-425.
107. M.-Y. Li, C.-H. Chen, Y. Shi and L.-J. Li, *Materials Today*, 2016, **19**, 322-335.
108. A. Pant, Z. Mutlu, D. Wickramaratne, H. Cai, R. K. Lake, C. Ozkan and S. Tongay, *Nanoscale*, 2016, **8**, 3870-3887.
109. H. Fang, C. Battaglia, C. Carraro, S. Nemsak, B. Ozdol, J. S. Kang, H. A. Bechtel, S. B. Desai, F. Kronast and A. A. Unal, *Proceedings of the National Academy of Sciences*, 2014, **111**, 6198-6202.
110. X. Hong, J. Kim, S.-F. Shi, Y. Zhang, C. Jin, Y. Sun, S. Tongay, J. Wu, Y. Zhang and F. Wang, *Nature nanotechnology*, 2014, **9**, 682-686.
111. M. M. Furchi, A. Pospischil, F. Libisch, J. Burgdörfer and T. Mueller, *Nano letters*, 2014, **14**, 4785-4791.
112. C. H. Lee, G. H. Lee, A. M. van der Zande, W. Chen, Y. Li, M. Han, X. Cui, G. Arefe, C. Nuckolls, T. F. Heinz, J. Guo, J. Hone and P. Kim, *Nat Nanotechnol*, 2014, **9**, 676-681.
113. Y. Deng, Z. Luo, N. J. Conrad, H. Liu, Y. Gong, S. Najmaei, P. M. Ajayan, J. Lou, X. Xu and P. D. Ye, *ACS nano*, 2014, **8**, 8292-8299.
114. M.-Y. Li, Y. Shi, C.-C. Cheng, L.-S. Lu, Y.-C. Lin, H.-L. Tang, M.-L. Tsai, C.-W. Chu, K.-H. Wei and J.-H. He, *Science*, 2015, **349**, 524-528.
115. D. Jariwala, V. K. Sangwan, C.-C. Wu, P. L. Prabhumirashi, M. L. Geier, T. J. Marks, L. J. Lauhon and M. C. Hersam, *Proceedings of the National Academy of Sciences*, 2013, **110**, 18076-18080.
116. S. Vélez, J. Island, M. Buscema, O. Txoperena, S. Parui, G. A. Steele, F. Casanova, H. S. van der Zant, A. Castellanos-Gomez and L. E. Hueso, *Nanoscale*, 2015, **7**, 15442-15449.
117. O. Lopez-Sanchez, E. Alarcon Llado, V. Koman, A. Fontcuberta i Morral, A. Radenovic and A. Kis, *Acs Nano*, 2014, **8**, 3042-3048.
118. H. Huang, Y. Huang, S. Wang, M. Zhu, H. Xie, L. Zhang, X. Zheng, Q. Xie, D. Niu and Y. Gao, *Crystals*, 2016, **6**, 113.
119. S. H. H. Shokouh, A. Pezeshki, A. Raza, S. Raza, H. S. Lee, S. W. Min, P. J. Jeon, J. M. Shin and S. Im, *Advanced Materials*, 2015, **27**, 150-156.
120. P. J. Jeon, Y. T. Lee, J. Y. Lim, J. S. Kim, D. K. Hwang and S. Im, *Nano letters*, 2016, **16**, 1293-1298.
121. S. H. Hosseini Shokouh, A. Pezeshki, S. R. A. Raza, K. Choi, S.-W. Min, P. J. Jeon, H. S. Lee and S. Im, *ACS nano*, 2014, **8**, 5174-5181.

122. M.-L. Tsai, S.-H. Su, J.-K. Chang, D.-S. Tsai, C.-H. Chen, C.-I. Wu, L.-J. Li, L.-J. Chen and J.-H. He, *ACS nano*, 2014, **8**, 8317-8322.
123. L. Hao, Y. Liu, W. Gao, Z. Han, Q. Xue, H. Zeng, Z. Wu, J. Zhu and W. Zhang, *Journal of applied physics*, 2015, **117**, 114502.
124. V. Dhyani and S. Das, *Scientific Reports*, 2017, **7**.
125. L. Wang, J. Jie, Z. Shao, Q. Zhang, X. Zhang, Y. Wang, Z. Sun and S. T. Lee, *Advanced Functional Materials*, 2015, **25**, 2910-2919.
126. P. Gehring, R. Urcuyo, D. L. Duong, M. Burghard and K. Kern, *Applied Physics Letters*, 2015, **106**, 233110.
127. P. Langevin, *Ann. Chim. Phys*, 1903, **28**, 122.
128. W. Shockley and W. Read Jr, *Physical review*, 1952, **87**, 835.
129. R. N. Hall, *Physical Review*, 1952, **87**, 387-387.
130. N. Greenham and P. Bobbert, *Physical Review B*, 2003, **68**, 245301.
131. M. M. Furchi, D. K. Polyushkin, A. Pospischil and T. Mueller, *Nano Lett*, 2014, **14**, 6165-6170.
132. J. O. Island, S. I. Blanter, M. Buscema, H. S. J. van der Zant and A. Castellanos-Gomez, *Nano Letters*, 2015, **15**, 7853-7858.
133. J. Wong, D. Jariwala, G. Tagliabue, K. Tat, A. R. Davoyan, M. C. Sherrott and H. A. Atwater, *ACS nano*, 2017, **11**, 7230-7240.
134. L. Ye, H. Li, Z. Chen and J. Xu, *ACS Photonics*, 2016, **3**, 692-699.
135. A. J. Molina-Mendoza, E. Giovanelli, W. S. Paz, M. A. Niño, J. O. Island, C. Evangeli, L. Aballe, M. Foerster, H. S. Van Der Zant and G. Rubio-Bollinger, *Nature Communications*, 2017, **8**.
136. A. Pezeshki, S. H. H. Shokouh, T. Nazari, K. Oh and S. Im, *Advanced Materials*, 2016, **28**, 3216-3222.
137. K. Zhang, T. Zhang, G. Cheng, T. Li, S. Wang, W. Wei, X. Zhou, W. Yu, Y. Sun and P. Wang, *ACS nano*, 2016, **10**, 3852-3858.
138. M. L. Tsai, M. Y. Li, J. R. D. Retamal, K. T. Lam, Y. C. Lin, K. Suenaga, L. J. Chen, G. Liang, L. J. Li and H. He Jr, *Advanced Materials*, 2017, **29**, 1701168.
139. Y. Ahn, A. Tsen, B. Kim, Y. W. Park and J. Park, *Nano letters*, 2007, **7**, 3320-3323.
140. F. Xia, T. Mueller, R. Golizadeh-Mojarad, M. Freitag, Y.-m. Lin, J. Tsang, V. Perebeinos and P. Avouris, *Nano Letters*, 2009, **9**, 1039-1044.
141. H. Yuan, X. Liu, F. Afshinmanesh, W. Li, G. Xu, J. Sun, B. Lian, A. G. Curto, G. Ye and Y. Hikita, *Nature nanotechnology*, 2015, **10**, 707-713.
142. G. Rao, M. Freitag, H.-Y. Chiu, R. S. Sundaram and P. Avouris, *ACS nano*, 2011, **5**, 5848-5854.
143. M. Tosun, D. Fu, S. B. Desai, C. Ko, J. S. Kang, D.-H. Lien, M. Najmzadeh, S. Tongay, J. Wu and A. Javey, *Scientific reports*, 2015, **5**, 10990.
144. S. Kallatt, G. Umesh, N. Bhat and K. Majumdar, *Nanoscale*, 2016, **8**, 15213-15222.
145. M. C. Lemme, F. H. Koppens, A. L. Falk, M. S. Rudner, H. Park, L. S. Levitov and C. M. Marcus, *Nano letters*, 2011, **11**, 4134-4137.
146. S. L. Howell, D. Jariwala, C.-C. Wu, K.-S. Chen, V. K. Sangwan, J. Kang, T. J. Marks, M. C. Hersam and L. J. Lauhon, *Nano Letters*, 2015, **15**, 2278-2284.
147. R. Sundaram, M. Engel, A. Lombardo, R. Krupke, A. Ferrari, P. Avouris and M. Steiner, *Nano letters*, 2013, **13**, 1416-1421.
148. L. Dobusch, S. Schuler, V. Perebeinos and T. Mueller, *Advanced Materials*, 2017, **29**, 1701304.
149. Y. Zhang, T. Oka, R. Suzuki, J. Ye and Y. Iwasa, *Science*, 2014, **344**, 725-728.
150. S. Jo, N. Ubrig, H. Berger, A. B. Kuzmenko and A. F. Morpurgo, *Nano letters*, 2014, **14**, 2019-2025.
151. E. Ponomarev, I. Gutiérrez-Lezama, N. Ubrig and A. F. Morpurgo, *Nano letters*, 2015, **15**, 8289-8294.
152. I. Gutiérrez-Lezama, B. A. Reddy, N. Ubrig and A. F. Morpurgo, *2D Materials*, 2016, **3**, 045016.
153. D. Li, R. Cheng, H. Zhou, C. Wang, A. Yin, Y. Chen, N. O. Weiss, Y. Huang and X. Duan, *Nature communications*, 2015, **6**.
154. R. Cheng, D. Li, H. Zhou, C. Wang, A. Yin, S. Jiang, Y. Liu, Y. Chen, Y. Huang and X. Duan, *Nano letters*, 2014, **14**, 5590-5597.
155. Y. Yang, N. Huo and J. Li, *Journal of Materials Chemistry C*, 2017, **5**, 7051-7056.
156. L. Wang, L. Huang, W. C. Tan, X. Feng, L. Chen and K.-W. Ang, *Advanced Electronic Materials*, 2018, **4**, 1700442.
157. Y. Jin, D. H. Keum, S. J. An, J. Kim, H. S. Lee and Y. H. Lee, *Advanced Materials*, 2015, **27**, 5534-5540.
158. Y. Liu, Y. Cai, G. Zhang, Y. W. Zhang and K. W. Ang, *Advanced Functional Materials*, 2017, **27**, 1604368.
159. J. Ahn, P. J. Jeon, S. R. A. Raza, A. Pezeshki, S.-W. Min, D. K. Hwang and S. Im, *2D Materials*, 2016, **3**, 045011.
160. Y. Gong, S. Lei, G. Ye, B. Li, Y. He, K. Keyshar, X. Zhang, Q. Wang, J. Lou and Z. Liu, *Nano letters*, 2015, **15**, 6135-6141.
161. N. Huo, J. Kang, Z. Wei, S. S. Li, J. Li and S. H. Wei, *Advanced Functional Materials*, 2014, **24**, 7025-7031.
162. F. Wang, Z. Wang, K. Xu, F. Wang, Q. Wang, Y. Huang, L. Yin and J. He, *Nano letters*, 2015, **15**, 7558-7566.
163. S. Chuang, R. Kapadia, H. Fang, T. Chia Chang, W.-C. Yen, Y.-L. Chueh and A. Javey, *Applied Physics Letters*, 2013, **102**, 242101.
164. T. Hong, B. Chamlagain, T. Wang, H.-J. Chuang, Z. Zhou and Y.-Q. Xu, *Nanoscale*, 2015, **7**, 18537-18541.
165. N. Flöry, A. Jain, P. Bharadwaj, M. Parzefall, T. Taniguchi, K. Watanabe and L. Novotny, *Applied Physics Letters*, 2015, **107**, 123106.
166. P. Chen, T. T. Zhang, J. Xiang, H. Yu, S. Wu, X. Lu, G. Wang, F. Wen, Z. Liu and R. Yang, *Nanoscale*, 2016, **8**, 3254-3258.

167. X. Wang, L. Huang, Y. Peng, N. Huo, K. Wu, C. Xia, Z. Wei, S. Tongay and J. Li, *Nano Research*, 2016, **9**, 507-516.
168. A.-J. Cho, S. D. Namgung, H. Kim and J.-Y. Kwon, *APL Materials*, 2017, **5**, 076101.
169. F. Yan, L. Zhao, A. Patanè, P. a. Hu, X. Wei, D. Zhang, Q. Lv, Q. Feng, C. Shen and K. Chang, *Nanotechnology*, 2017.
170. W. C. Yap, Z. Yang, M. Mehboudi, J.-A. Yan, S. Barraza-Lopez and W. Zhu, *Nano Research*, 2017, **11**, 420-430.
171. C. Li, X. Yan, X. Song, W. Bao, S. Ding, D. W. Zhang and P. Zhou, *Nanotechnology*, 2017, **28**, 415201.
172. S. Yang, M. Wu, B. Wang, L.-D. Zhao, L. Huang, C. Jiang and S.-H. Wei, *ACS applied materials & interfaces*, 2017, **9**, 42149-42155.
173. A. S. Aji, M. Izumoto, K. Suenaga, K. Yamamoto, H. Nakashima and H. Ago, *Physical Chemistry Chemical Physics*, 2018, **20**, 889-897.
174. T. Yang, B. Zheng, Z. Wang, T. Xu, C. Pan, J. Zou, X. Zhang, Z. Qi, H. Liu and Y. Feng, *Nature communications*, 2017, **8**, 1906.
175. K. Zhang, M. Peng, W. Wu, J. Guo, G. Gao, Y. Liu, J. Kou, R. Wen, Y. Lei and A. Yu, *Materials Horizons*, 2017, **4**, 274-280.
176. F. Liu, W. L. Chow, X. He, P. Hu, S. Zheng, X. Wang, J. Zhou, Q. Fu, W. Fu and P. Yu, *Advanced Functional Materials*, 2015, **25**, 5865-5871.
177. D. Jariwala, S. L. Howell, K.-S. Chen, J. Kang, V. K. Sangwan, S. A. Filippone, R. Turrisi, T. J. Marks, L. J. Lauhon and M. C. Hersam, *Nano letters*, 2015, **16**, 497-503.
178. D. He, Y. Pan, H. Nan, S. Gu, Z. Yang, B. Wu, X. Luo, B. Xu, Y. Zhang and Y. Li, *Applied Physics Letters*, 2015, **107**, 183103.
179. D. Sarkar, X. Xie, W. Liu, W. Cao, J. Kang, Y. Gong, S. Kraemer, P. M. Ajayan and K. Banerjee, *Nature*, 2015, **526**, 91-95.
180. Y. Yu, P. W. Fong, S. Wang and C. Surya, *Scientific reports*, 2016, **6**.
181. C. Yim, M. O'brien, N. McEvoy, S. Riazimehr, H. Schäfer-Eberwein, A. Bablich, R. Pawar, G. Iannaccone, C. Downing and G. Fiori, *Scientific reports*, 2014, **4**, 5458.
182. Z. Wang, M. Li, L. Yang, Z. Zhang and X. P. Gao, *Nano Research*, 2017, **10**, 1872-1879.
183. Y. Niu, R. Frisenda, S. A. Svatek, G. Orfila, F. Gallego, P. Gant, N. Agraït, C. Leon, A. Rivera-Calzada and D. P. De Lara, *2D Materials*, 2017, **4**, 034002.
184. K. Kang, S. Xie, L. Huang, Y. Han, P. Y. Huang, K. F. Mak, C.-J. Kim, D. Muller and J. Park, *Nature*, 2015, **520**, 656.
185. Y. Huang, J. Qiao, K. He, S. Bliznakov, E. Sutter, X. Chen, D. Luo, F. Meng, D. Su and J. Decker, *Chemistry of Materials*, 2016, **28**, 8330-8339.
186. F. Alsaffar, S. Alodan, A. Alrasheed, A. Alhussain, N. Alrubaiq, A. Abbas and M. R. Amer, *Scientific Reports*, 2017, **7**, 44540.
187. R. A. Doganov, E. C. O'Farrell, S. P. Koenig, Y. Yeo, A. Ziletti, A. Carvalho, D. K. Campbell, D. F. Coker, K. Watanabe and T. Taniguchi, *Nature communications*, 2015, **6**.
188. A. Avsar, I. J. Vera-Marun, J. Y. Tan, K. Watanabe, T. Taniguchi, A. H. Castro Neto and B. Ozyilmaz, *ACS nano*, 2015, **9**, 4138-4145.
189. Y. Cao, A. Mishchenko, G. Yu, E. Khestanova, A. Rooney, E. Prestat, A. Kretinin, P. Blake, M. Shalom and C. Woods, *Nano letters*, 2015, **15**, 4914-4921.
190. M. Velický, P. S. Toth, A. M. Rakowski, A. P. Rooney, A. Kozikov, C. R. Woods, A. Mishchenko, L. Fumagalli, J. Yin and V. Zólyomi, *Nature Communications*, 2017, **8**, 14410.
191. G. Prando, *Nature Nanotechnology*, 2017, **12**, 191-191.
192. Y. Jiang, J. Mao, D. Moldovan, M. R. Masir, G. Li, K. Watanabe, T. Taniguchi, F. M. Peeters and E. Y. Andrei, *Nature Nanotechnology*, 2017, **12**, 1045.
193. R. Heinisch, F. Bronold and H. Fehske, *Physical Review B*, 2013, **87**, 155409.
194. M. Huang, S. Li, Z. Zhang, X. Xiong, X. Li and Y. Wu, *Nature nanotechnology*, 2017, **12**, 1148-1154.
195. D. Akinwande, N. Petrone and J. Hone, *Nature communications*, 2014, **5**, 5678.
196. G. Fiori, F. Bonaccorso, G. Iannaccone, T. Palacios, D. Neumaier, A. Seabaugh, S. K. Banerjee and L. Colombo, *Nature Nanotechnology*, 2014, **9**, 768-779.
197. F. Xia, H. Wang, D. Xiao, M. Dubey and A. Ramasubramaniam, *Nature Photonics*, 2014, **8**, 899.